\documentclass[useAMS,usenatbib]{mn2e}
\usepackage[centertags]{amsmath}
\usepackage{amsfonts}
\usepackage{amssymb}
\usepackage{newlfont}
\usepackage{textcomp}
\usepackage{times}
\usepackage{graphicx}
\usepackage{subfigure}
\usepackage{psfrag}
\usepackage{amsmath}
\usepackage[at]{easylist}
\usepackage{bm}
\usepackage{color}

\usepackage{stmaryrd}

\def\reff@jnl#1{{\rm#1\/}}
\def\apj{\reff@jnl{ApJ}}       
\def\apjs{\reff@jnl{ApJS}}     
\def\aaps{\reff@jnl{A\&AS}}    
\def\mnras{\reff@jnl{MNRAS}}   
\def\prd{\reff@jnl{Phys.\ Rev.\ D}}    

\newcommand{\beq}{\begin{equation}}
\newcommand{\eeq}{\end{equation}}

\newcommand{\be}{\begin{equation}}
\newcommand{\ee}{\end{equation}}

{\begin{enumerate}\setlength{\itemsep}{0mm}}%
{\end{enumerate}}
{\begin{enumerate}\setlength{\itemsep}{0mm}}%
{\end{enumerate}}

\newcommand{\planck}{{\it Planck }}

\def\vect#1{{\bmath{#1}}}

\title[Bayesian validation]{Validation of Bayesian posterior distributions using a multidimensional Kolmogorov--Smirnov test}
\author[D.L.~Harrison et al.]
{Diana Harrison,$^{1, 2}$\thanks{email: dlh@ast.cam.ac.uk}
 David Sutton,$^{1,2}$\thanks{email: sutton@ast.cam.ac.uk}
 Pedro Carvalho,$^{1,2}$\thanks{email: carvalho@mrao.cam.ac.uk}
 Michael Hobson$^{3}$\thanks{email: mph@mrao.cam.ac.uk}
\\
\\
  $^1$ Kavli Institute for Cosmology Cambridge, Madingley Road, Cambridge, CB3 0HA, U.K.\\
  $^2$ Institute of Astronomy, Madingley Road, Cambridge, CB3 0HA, U.K.\\
  $^3$ Astrophysics Group, Cavendish Laboratory, J.J.~Thomson Avenue,
Cambridge, CB3 0HE, U.K.\\
}

\date{Accepted ---. Received ---; in original form \today}
\pagerange{\pageref{firstpage}--\pageref{lastpage}}
\pubyear{2014}

\begin{document}

\maketitle
\label{firstpage}

\begin{abstract}
We extend the Kolmogorov--Smirnov (K-S) test to multiple dimensions by
suggesting a $\mathbb{R}^n \rightarrow [0,1]$ mapping based on the
probability content of the highest probability density region of the
reference distribution under consideration; this mapping reduces the
problem back to the one-dimensional case to which the standard K-S
test may be applied. The universal character of this mapping also
allows us to introduce a simple, yet general, method for the
validation of Bayesian posterior distributions of any dimensionality.
This new approach goes beyond validating software implementations; it
provides a sensitive test for all assumptions, explicit or implicit,
that underlie the inference. In particular, the method assesses
whether the inferred posterior distribution is a truthful
representation of the actual constraints on the model parameters.  We
illustrate our multidimensional K-S test by applying it to a simple
two-dimensional Gaussian toy problem, and demonstrate our method for
posterior validation in the real-world astrophysical application of
estimating the physical parameters of galaxy clusters parameters from
their Sunyaev--Zel'dovich effect in microwave background data.  In the
latter example, we show that the method can validate the entire Bayesian
inference process across a varied population of objects for which the
derived posteriors are different in each case.
\end{abstract}
\section{Introduction}
\label{sect:Introduction}
The Kolmogorov--Smirnov (K-S) test and its close variants are the most
widely-accepted methods for testing whether a set of continuous sample
data is drawn from a given probability distribution function $f(x)$
\citep[see, e.g.,][ch. 14.3.3]{NumericalRecipes2007}.  The test is
based on the computation of the maximum absolute distance $D$ between
the cumulative distribution functions (CDFs) of the data and of
$f(x)$.  In particular, it has the advantage that, in the `null
hypothesis' that the data are indeed drawn from $f(x)$, the
distribution of $D$ is both independent of the form of $f(x)$
(i.e. the test is `distribution-free') and can be calculated, at least
to a good approximation. Moreover, the test is invariant under
reparameterizations of $x$. In its standard form, however, the K-S
test is restricted to one-dimensional data, since a CDF can only be
defined uniquely in the univariate case.

Considerable effort has been made to extend the K-S test to
$n$-dimensional data sets, much of it building on the pioneering work
of \cite{Peacock_KS}, but this has proved very challenging precisely
because a CDF is not well-defined in more than one dimension.  Working
initially in two dimensions, Peacock's original insight was to replace
the notion of a CDF with the integrated probability in each of the
four natural quadrants around any given data point $(x_i,y_i)$, and
define the distance measure $D$ as the maximum absolute difference
(ranging both over data points and quadrants) of the corresponding
integrated probabilities for the data and the theoretical distribution
$f(x,y)$. This basic idea can, in principle, be extended
straightforwardly to higher dimensions \citep{Gosset1987}, but in
practice suffers from an exponential growth in complexity, since the
number of independent integrated probabilities about any given data
point is $2^n-1$ in $n$-dimensions, although \cite{Franceschini1987}
suggest an algorithm with better scaling. Perhaps a more notable
deficit is that, in the null hypothesis, the distribution of $D$ is
{\em not} independent of the form of $f(x,y)$, although
\cite{NumericalRecipes2007} reassure us that extensive Monte Carlo
simulations show the distribution of the two-dimensional $D$ to be
very nearly identical for even quite different distributions, provided
they have the same correlation coefficient \citep{Franceschini1987} .
\cite{Lopes2008} contains a review that ranks the performance of a
range of different multidimensional K-S tests that are variants of
Peacock's original proposal. A completely different approach to
extending the K-S test to multiple dimensions was advanced by
\cite{Justel1997} and employs the Rosenblatt transformation, which
maps any distribution into the unit hyper-cube
\citep{rosenblatt1952}. Despite this approach being formally sound,
the authors report insurmountable difficulties when extending the test
to more than two dimensions.

In this paper, we present a new proposal for extending the K-S test to
multiple dimensions, which is free from the weaknesses discussed above
and scales straightforwardly with the dimensionality of the data-set.
Our approach is to introduce the $\mathbb{R}^n \rightarrow [0,1]$
mapping from any given (data) point $\bmath{x}_i$ in the
$n$-dimensional parameter space to the probability mass $\zeta$ of the
putative theoretical distribution $f(\bmath{x})$ contained within the
highest probability density (HPD) region having $\bmath{x}_i$ on its
boundary.  This mapping has the convenient property that under the
null hypothesis that the data are drawn from $f(\bmath{x})$, the
probability mass $\zeta$ is uniformly distributed in the range
$[0,1]$, independently of the form of $f(\bmath{x})$. The set of values
$\{\zeta_i\}$ corresponding to the data points
$\{\bmath{x}_i\}$ can then be compared with the uniform distribution
in a standard one-dimensional K-S test (or one of its variants).

The ability to test the hypothesis that a set of data samples are
drawn from some general $n$-dimensional probability distribution
$f(\bmath{x})$ has an interesting application in the validation of
Bayesian inference analyses (indeed this application provided our
original motivation for seeking to extend the K-S test to multiple
dimensions).  Bayesian methods are now pervasive across all branches
of science and engineering, from cognitive neuroscience
\citep{Doya:2007} and machine learning
\citep{Bishop:2006:PRM:1162264}, to spam filtering
\citep{Sahami98abayesian} and geographic profiling
(\citealt{geographic_profiling},
\citealt{LandminesBayesianDetection}).  In precision cosmology,
Bayesian inference is the main tool for setting constraints on
cosmological parameters \citep{PlanckParameters,WMAP9Parameters}, but
very few attempts have been made to assess whether the derived
posterior probability distributions are a truthful representation of
the actual parameter constraints one can infer from the data in the
context of a given physical model.  This lack of validation has been
highlighted by several authors, with the strong dependence of the
inference results on the priors being of particular
concern \citep{GeorgeBayes,LindMiq}. There have been attempts to
address this issue, ranging from the approach of
\citet{Cook06validationof}, which was designed with software
validation solely in mind, to a method based on the inverse
probability integral transform (Smirnov transform) applied to
posterior distributions, that extends to spaces of higher
dimensionality via marginalisation \citep{FnlVal}. Also, validation of
the Bayesian source-finding algorithm of \cite{PwSII} was performed in
\citet{PlanckResultsSZ}, but only point estimates deduced from the
posterior distributions were actually verified.
Our method for addressing this problem is based on our applying our
multidimensional extension of the K-S test to sets of Monte-Carlo
simulations of the data and the posterior distributions derived
therefrom. In particular, it can take advantage of the fact that one
may typically generate simulations that are of greater sophistication
and realism than may be modelled in the inference process, and thus
allows for a more thorough validation of the inference than has been
possible with the methods developed previously.  In particular, our
validation procedure enables us to test all the assumptions made
(explicitly or implicitly) in the inference process, such as the
statistical description of the data, model assumptions and
approximations, as well as the software implementation of the
analysis. Moreover, we consider the full posterior distribution,
regardless of its dimensionality and form, without the need to resort
to marginalization, and thereby keeping intact its $n$-dimensional
character.

This paper is organised as follows. Section~\ref{sect:MathFrame}
provides the mathematical background for our extension of the K-S test
to multiple dimensions and Section~\ref{sec:Validation} describes its
application to the validation of Bayesian inference analyses.  In
Section~\ref{sect:ToyModel} we apply our multidimensional K-S test to
a simple toy problem, and in Section~\ref{sect:PlanckSZcase} we illustrate
our Bayesian inference validation procedure by applying it to the real
astronomical example of detecting and characterising galaxy clusters
in observations of the microwave sky through their Sunyaev--Zel'dovich
effect. We conclude in Section~\ref{sect:Conclusions}.

\section{Multidimensional K-S test}
\label{sect:MathFrame}
\subsection{Classical one-dimensional K-S test and its variants}
\label{sect:ndimKS}
The classical one-dimensional K-S test is based on the distance
measure $D$ defined by
\begin{equation}
D = \underset{x}{\mbox{max}} ~ |S_N(x) - F(x)|,
\label{eqn:ksdistance}
\end{equation}
where $F(x) = \int_{-\infty}^{x} f(u)\,du$ is the cumulative
distribution function (CDF) of the putative theoretical probability
distribution $f(x)$ and $S_N(x)$ is the empirical CDF of the data-set
$\{x_1,x_2,\ldots x_N\}$ that one wishes to test. As mentioned above,
under the null hypothesis that the data are drawn from $f(x)$, the
probability distribution of $D$ is independent of the form of $f(x)$.
To a very good approximation, the probability of $D$ being larger than
the observed value $D_\mathrm{obs}$ is
\begin{equation}
{\Pr}(D > D_{\mathrm{obs}}) = \mathcal{Q}_\mathrm{KS}([\sqrt{N} + 0.12 +
    0.11/\sqrt{N}] D_{\mathrm{obs}}),
\label{eqn:ksdistribution}
\end{equation}
where the Kolmogorov distribution $Q_\mathrm{KS}$ is given
by\footnote{An efficient implementation for evaluating $Q_\mathrm{KS}$
  may be found in \cite{NumericalRecipes2007}.}
\begin{equation}
\mathcal{Q}_\mathrm{KS}(u) = \sum_{n=1}^{\infty} (-1)^{n-1}
\exp(-2n^2u^2).
\end{equation}
Moreover, the maximum distance $D$,
and hence the K-S test itself, is invariant under reparameterization
of $x$.

It is well known, however, that the K-S test is not equally sensitive
over the entire range of the variable $x$, since under the null
hypothesis the variance of $|S_N(x)-F(x)|$ is proportional to $f(x) [1
  - f(x)]$. Consequently, the K-S test is most sensitive around the
median point, where $f(x)=0.5$, and less sensitive in the tails of the
distribution. This makes the K-S test particularly sensitive to
`shifts' (in particular differences in the median), but less sensitive
to differences in the dispersion, particularly if the median is
similar. Two popular variants of the K-S test seek to remedy this
shortcoming in different ways. \cite{anderson1952} suggest correcting
for the effect by `stabilizing' the statistic $D$ using a weighting
scheme based on the standard deviation of the null hypothesis,
although the probability distribution of the resulting `distance'
statistic does not have a simple form. A more direct approach,
proposed by \cite{Kuiper1960}, is to wrap the $x$-axis around into a
circle (by identifying the points at $x=\pm\infty$) and define a new
`distance' statistic that is invariant under all reparameterizations
on the circle. This guarantees equal sensitivity at all values of $x$
and there exists a simple formula for the asymptotic distribution of
the statistic. Clearly, this variant is ideally suited for
distributions originally defined on a circle.

\subsection{Highest probability density regions}
\label{sect:Hpds}

To extend the K-S test (and its variants) to multiple dimensions, one
must clearly replace the use of the cumulative distribution function
(CDF), since this is not uniquely defined in more than one
dimension. In the univariate case, the CDF simply maps the value of
the random variable $x$ to the integrated probability under its PDF up
to that $x$-value.  This suggests that to accommodate multiple
dimensions one should seek an alternative mapping, that is
well-defined in any dimensionality, from the multidimensional random
variable $\bmath{x}$ to an associated integrated probability under
$f(\bmath{x})$. We therefore choose here to use the probability mass
$\zeta$ contained within the highest probability density (HPD) region
having $\bmath{x}$ on its boundary.

\begin{figure}
  \begin{center}
    \leavevmode
       \includegraphics[width=0.5\textwidth]{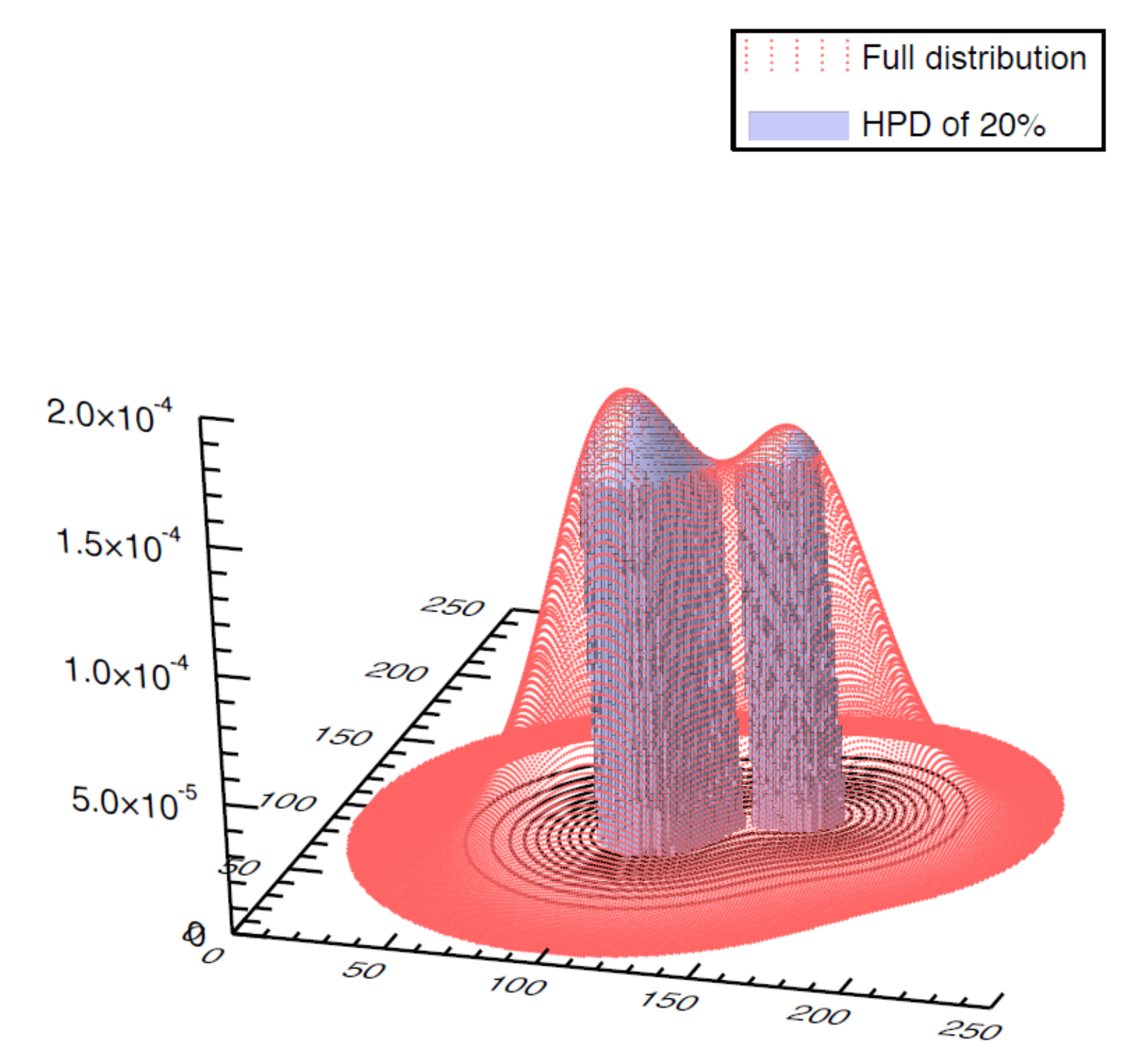}
    \caption{Illustration of the definition of the highest probability
      density region, HPD, in a 2-dimensional case. The unique
      HPD that encloses $20\%$ of the probability is shown.  All
      the values that are contained within the HPD have higher
      probability densities than those that lie outside
      this region. }
    \label{fig:HPD_definition}
  \end{center}
\end{figure}

HPDs are discussed by \citet{BoxTiao} as a general method for
compressing the information contained within a probability
distribution $f(\bmath{x})$ of arbitrary dimensional and form.  Each
HPD is uniquely defined by the amount of probability $\zeta$ it
encloses and is constructed such that there exists no probability
density value outside the HPD that is greater than any value contained
within it.  More formally, $\mbox{HPD}_{\zeta}$ is the
region in $\bmath{x}$-space defined by $\Pr(\bmath{x} \in
\mbox{HPD}_\zeta)=\zeta$ and the requirement that if
$\bmath{x}_1\in\mbox{HPD}_\zeta$ and
$\bmath{x}_2\not\in\mbox{HPD}_\zeta$ then $f(\bmath{x}_1) \ge
f(\bmath{x}_2)$.  Figure~\ref{fig:HPD_definition} provides a graphical
illustration of the HPD definition in two dimensions.

Some properties of HPD regions are summarised in
Appendix~\ref{sec:HPDprops}. In particular, we show that
$\mbox{HPD}_{\zeta}$ is the region with smallest volume that encloses
a total probability $\zeta$; thus the HPD provides an elegant way of
communicating inference results, since it corresponds to the smallest
region of parameter space having a given level of (un)certainty
$\zeta$.

We may use the notion of the HPD to define a many-to-one $\mathbb{R}^n
\rightarrow [0,1]$ mapping between any point $\bmath{x}$ in the
$n$-dimensional parameter space and the probability content $\zeta$ of
the HPD whose boundary passes through $\bmath{x}$, namely
\begin{equation}
\zeta(\bmath{x}) = \int_{f(\bmath{u}) \geq f(\bmath{x})}^{}
f(\bmath{u}) ~ d^{n}\bmath{u}.
\label{eq:HPD_varDef}
\end{equation}
Thus, in general, each point $\bmath{x}$ in the parameter space maps
onto one and only one value of $\zeta$, but, obviously, more than one
point $\bmath{x}$ might map onto the same $\zeta$-value. In one
dimension, this contrasts with the more desirable one-to-one mapping
between the value of the random variable $x$ and its CDF $F(x)$, but
the latter cannot be generalised to multiple dimensions.

It is easy to show that, under the null hypothesis that a set of data
points $\{\bmath{x}_i\}$ are drawn from $f(\bmath{x})$, then the
corresponding set of values $\{\zeta_i\}$ are drawn from the standard
uniform distribution $U(0,1)$, {\em independently} of the form of
$f(\bmath{x})$. This may be seen by differentiating
(\ref{eq:HPD_varDef}) to obtain
\begin{equation}
d\zeta = f(\bmath{x})\,d^n \bmath{x},
\label{eq:deltaZ_Empirical}
\end{equation}
which simply shows that if we add an element of volume $d^n \bmath{x}$ to
the HPD, its probability content increases by $f(\bmath{x})\,d^n
\bmath{x}$. Comparing this equation with the standard result for a
change of variables $\bmath{x} \to \zeta$ in probability
distributions, namely
\begin{equation}
g(\zeta)\,d\zeta = f(\bmath{x})\,d^n \bmath{x},
\label{eq:VarChangingGen}
\end{equation}
one immediately sees that $g(\zeta)=1$ (over its allowed range $[0,1]$)
and hence $\zeta$ follows the standard uniform distribution. A more
rigorous proof is given in Appendix \ref{sec:HPDprops}. As we
will see, this simple result allows for the straightforward extension
of the K-S test to multiple dimensions and is central to its
subsequent use in the validation of Bayesian inference analyses.

\subsection{Pathological cases}
\label{sec:pathological}

Before continuing with the mathematical development, it is necessary
to discuss some pathological cases that require further
consideration.  Such cases occur when there exist perfect `plateau'
regions of the parameter space over which $f(\vect{x})$ takes
precisely the same value.

Let us begin by rewriting the mapping (\ref{eq:HPD_varDef}) as
\begin{equation}
\zeta(\bmath{x}) =
\int_{f(\bmath{u}) > f(\bmath{x})}^{}
f(\bmath{u}) ~ d^{n}\bmath{u} +
\int_{f(\bmath{u}) = f(\bmath{x})}^{}
f(\bmath{u}) ~ d^{n}\bmath{u}.
\label{eq:HPD_varDef2}
\end{equation}
If the point $\vect{x}$ lies in a perfect plateau region $\mathcal{R}$
(which may, in general, consist of multiple disjoint subregions)
over which $f(\vect{x})$ takes some constant value $f_\ast$, then the
second integral on the RHS of (\ref{eq:HPD_varDef2}) has the {\em
  finite} value $V(\mathcal{R})f_\ast$, where $V(\mathcal{R})$ is the
$n$-dimensional volume of the region $\mathcal{R}$. Thus,
(\ref{eq:HPD_varDef2}) maps {\em every} point in $\mathcal{R}$ to the
{\em same} value of $\zeta$.

As a result, even if a set of data points $\{\vect{x}_i\}$ are drawn
from $f(\vect{x})$, the corresponding values $\{\zeta_i\}$ will {\em
  not} be drawn from the standard uniform distribution $U(0,1)$.  The
extreme example is when $f(\vect{x})$ is the uniform distribution over
the entire parameter space, in which case the mapping
(\ref{eq:HPD_varDef2}) yields the value $\zeta_i = 1$ for every point
$\vect{x}_i$ (in this case the entire contribution to $\zeta$ comes
from the second integral on the RHS of (\ref{eq:HPD_varDef2}), since
the first one vanishes).

Fortunately, we can accommodate such pathological cases
  by using the fact that data points drawn from $f(\vect{x})$ are
  uniformly distributed within the perfect plateau regions. For such a
  set of data points, which exhibit complete spatial randomness, it is
  straightforward to show that, in $N$ dimensions and ignoring edge
  effects, the CDF of the nearest neighbour Euclidean distance $d$ is
  $G(d)=1-\exp[-\rho V_N(d)]$, where
  $V_N(d)=\pi^{n/2}d^n/\Gamma(\frac{n}{2}+1)$ is the `volume' of the
  $N$-ball of radius $d$ (such that $V_2(d)=\pi d^2$, $V_3(d) =
  \frac{4}{3}\pi d^3$, etc.), and $\rho$ is the expected number of
  data points per unit `volume'. Similarly, again ignoring edge
  effects, the CDF $F(s)$ of the distance $s$ from a randomly chosen
  location (not the location of a data point) in a perfect plateau
  region and the nearest data point has the {\em same} functional form
  as $G(d)$. Most importantly, as described by \cite{dixon12}, one may
  remove edge effects by combining $G(d)$ and $F(s)$. In particular,
  provided the data points exhibit complete spatial randomness, the
  quantity $J(d) \equiv [1-G(d)]/[1-F(d)]$ will equal unity for all
  $d$ and does not suffer from edge effects.

We make use of the $J(d)$
construction as follows.  For each of the data points $\bmath{x}_i$
$(i=1,2,\ldots,n)$ in the perfect plateau region ${\cal R}$, one first
calculates the distance $d_i$ to its nearest neighbouring data point
and evaluates $G(d_i)$ as given above. One then draws $n$ random
locations $\bmath{r}_j$ $(j=1,2,\ldots,n)$ in ${\cal R}$, calculates
the distance $s_j$ from each random location to its nearest data point
and evaluates $F(s_j)$ as given above.  Provided the data points are
completely spatially random and ignoring edge-effects, the sets of
values $\{G(d_i)\}$ and $\{F(s_j)\}$ are, respectively, distributed
according to the standard uniform distribution $U(0,1)$.
Consequently, the sets of values $\{1-G(d_i)\}$ and $\{1-F(s_j)\}$ are
similarly distributed. Hence, even in the presence of edge effects,
the quantities $J_i = [1-G(d_i)]/[1-F(s_i)]$ have the CDF $P(J) = J/2$
if $0 < J \leq 1$ and $(2-J^{-1})/2$ if $J > 1$. Thus, for the $i$th
data point, one may replace the expression (\ref{eq:HPD_varDef2}) by
\begin{equation}
\zeta(\bmath{x}_i) =
\int_{f(\bmath{u}) > f(\bmath{x}_i)}^{}
f(\bmath{u}) ~ d^{n}\bmath{u} +
f_\ast V({\cal R}) P(J_i),
\end{equation}
which is uniformly distributed, thereby recovering the behaviour in
the non-pathological case.

\subsection{Testing the null hypothesis}

It is now a simple matter to define our multidimensional extension to
the K-S test. Given a set of data points $\{\vect{x}_i\}$,
we may test the null hypothesis that they are drawn from the reference
distribution $f(\vect{x})$ as follows. First, we apply the mapping
(\ref{eq:HPD_varDef}) (or the mapping (\ref{eq:HPD_varDef2}) in
pathological cases) to obtain the corresponding set of values
$\{\zeta_i\}$. Second, we simply use the standard one-dimensional
K-S test (or one of its variants) to test the hypothesis that the
values $\{\zeta_i\}$ are drawn from the standard normal distribution
$U(0,1)$.

In this way, one inherits the advantages of the standard
one-dimensional K-S test, while being able to accommodate multiple
dimensions. In particular, since under the null hypothesis the set of
values $\{\zeta_i\}$ are uniformly distributed independently of the
form of $f(\vect{x})$, our approach retains the property enjoyed by
the one-dimensional K-S test of being distribution-free; this
contrasts with the method proposed by \cite{Peacock_KS}.

We note that, once the $\{\zeta_i\}$ values have been obtained, one
could test whether they are drawn from a uniform distribution using
any standard one-dimensional test. For example, if one wished, one
could bin the $\{\zeta_i\}$ values and use the chi-square test
\citep[see, e.g.,][]{NumericalRecipes2007}, although there is, of
course, an arbitrariness in choosing the bins and an inevitable loss
of information. For continuous one-dimensional data, such as the
$\{\zeta_i\}$ values, the most generally accepted test is the K-S test
(or one of its variants).

We also note that our approach may be applied even in the absence of
an analytical form for $f(\bmath{x})$, provided one has some
(numerical) means for evaluating it at any given point $\bmath{x}$. An
important example is when one has only a set of samples known to be
drawn from $f(\bmath{x})$.  In this case, one might evaluate
$f(\bmath{x})$ at any given point using some kernel estimation method
(say). If, however, one also knows the value of $f(\bmath{x})$ at each
sample point, a more elegant approach is to estimate the probability
mass $\zeta_i$ corresponding to each original (data) point
$\bmath{x}_i$ by calculating the fraction of samples for which
$f(\bmath{x}) \ge f(\bmath{x}_i)$.

\subsection{Performing multiple simultaneous tests}
\label{sec:multipletests}

The advantageous property that, under the null hypothesis, the derived
set of values $\{\zeta_i\}$ are drawn from the standard uniform
distribution $U(0,1)$, independently of the form of $f(\bmath{x})$,
allows us straightforwardly to perform multiple simultaneous tests as
follows.

Suppose that one has $M$ independent $n$-dimensional data-sets
$\{\vect{x}^{(m)}_i\}$, where $m = 1,2,\ldots, M$ labels data-sets and
$i=1,2,\ldots,N_m$ labels the data points in each set. Suppose further
that we wish to perform a simultaneous test of the null hypothesis
that each data-set $\{\vect{x}^{(m)}_i\}$ is drawn from the
corresponding reference distribution $f_m(\vect{x})$. This may be
achieved by first performing a transformation analogous to
(\ref{eq:HPD_varDef}) for each data set separately, namely
\begin{equation}
\zeta^{(m)}(\bmath{x}) = \int_{f_m(\bmath{u}) \geq f_m(\bmath{x})}^{}
f_m(\bmath{u}) ~ d^{n}\bmath{u}.
\label{eq:multiHPD_varDef}
\end{equation}
Under the null hypothesis, the resulting {\em combined} set of values
$\{\zeta^{(1)}_i,\zeta^{(2)}_i,\ldots,\zeta^{(M)}_i\}$ will again be
drawn from the standard uniform distribution, which can be tested, as
previously, using a single one-dimensional K-S test.

As an extreme example, this approach may even be used when one has
just a single data point drawn from each reference distribution
$f_m(\vect{x})$, in which case it would be impossible to perform an
individual test for each data point. We will make use of this
capability in applying our method to the validation of Bayesian
inference analyses in Section~\ref{sec:Validation}.

\subsection{Reparameterization invariance}
\label{sect:BayReParam}

A common criticism of HPD regions is that they are not invariant under
smooth reparameterizations \citep[][sec. 5.7]{lehmann2006testing}.  In
other words, if one performs a smooth change of variables $\vect{x}
\to \btheta = \btheta(\vect{x})$, then in general
$\mbox{HPD}_\zeta(\btheta) \neq \btheta(\mbox{HPD}_\zeta(\vect{x}))$,
although the image $\btheta(\mbox{HPD}_\zeta(\vect{x}))$ in
$\btheta$-space of the $\vect{x}$-space HPD region will still contain a
probability mass $\zeta$. This deficiency can be problematic for using
HPD regions to summarise the properties of probability density
functions, but it causes less difficulty here, since our only use of
the HPD region is to define the mapping (\ref{eq:HPD_varDef}).

Suppose that a set of data points $\{\vect{x}_i\}$ are drawn from
$f(\vect{x})$ and are transformed into the set $\{\btheta_i\}$, which
will be drawn from the corresponding PDF $h(\btheta)$ in the new
variables. Applying an analogous mapping to (\ref{eq:HPD_varDef}) to
the new data set $\{\btheta_i\}$ will, in general, lead to a different
set of values $\{\zeta_i\}$ to those derived from the original data
$\{\vect{x}_i\}$. One may visualise this straightforwardly by
considering any one of the data points $\vect{x}_i$, for which
$f(\vect{x}_i)=f_i$ (say), and the corresponding HPD region having
$\vect{x}_i$ on its boundary, which is defined simply by the contour
$f(\vect{x})=f_i$.  Performing a smooth mapping $\vect{x} \to \btheta
= \btheta(\vect{x})$, the data point $\vect{x}_i$ will be transformed
into $\btheta_i$, which will clearly lie within the image in
$\btheta$-space of the $\vect{x}$-space HPD, although not necessarily on its
boundary. This image region will contain the same probability mass as
the original $x$-space HPD, but it will in general not coincide with
the HPD in $\btheta$-space having $\btheta_i$ on its boundary, which
may contain a different probability mass. Nonetheless, subject to the
caveats above regarding pathological cases, the values $\{\zeta_i\}$
derived from the transformed data $\{\btheta_i\}$ will still be
uniformly distributed in the interval $[0,1]$, since this property
does not depend on the form of the PDF under consideration.

This does not mean, however, that the significance of our
multidimensional K-S test is invariant under
reparameterizations. Under the alternative hypothesis that the data
are not drawn from the reference distribution, the transformation
$\{\vect{x}_i\} \to \{\btheta_i\}$ may result in new derived values
$\{\zeta_i\}$ that differ more significantly from the standard uniform
distribution than the original set.

There exists the scope for adapting our approach, so that the
significance of the test is reparameterization invariant, by replacing
the HPD with an alternative mapping from the $n$-dimensional parameter
space $\vect{x}$ to some other measure of integrated probability under
$f(\vect{x})$. One such possibility is to use the `intrinsic credible
interval' (ICI) proposed by \cite{bernardo2005} , which has numerous desirable properties,
  including invariance under both reparameterization and
  marginalisation. We have not pursued this approach here, however,
  since the definition of the ICI is rather complicated and its
  calculation often requires numerical integration.  It also does not
  follow that the resulting mapping is distribution-free. We therefore
  leave this as a topic for future work.

In the context of the validation of Bayesian inference analyses,
however, in Section~\ref{sec:validrepar} we will discuss a method by
which one may retain the use of the HPD and ensure that the values of
$\{\zeta_i\}$ derived from the original and transformed data sets,
respectively, are in fact identical under smooth
reparameterizations. Thus, the significance of our resulting
multidimensional K-S test will be similarly invariant in this case.

\subsection{Iso-probability degeneracy}

We conclude this section, by noting that, even when our test gives a
`clean bill of health' to a set of data points, it unfortunately does
not guarantee immediately that those points were drawn from the
reference distribution. The basic test described above provides only a
\emph{necessary} condition; for it to be {\em sufficient} (at least in
the limit of a large number of data points) requires further
refinement, which we now discuss.

To understand this limitation of our basic test, we note, as mentioned
previously, that for the mapping (\ref{eq:HPD_varDef}) more than one
point $\bmath{x}$ in the parameter space may map onto the same
$\zeta$-value. Indeed, given a point $\bmath{x}$, the boundary of the
corresponding HPD is an iso-probability surface, and any other point
lying on this surface will naturally be mapped onto the same
$\zeta$-value. This constitutes the iso-probability degeneracy that is
inherent to the mapping (\ref{eq:HPD_varDef}). Thus, for example, if
one drew a set of data points $\{\bmath{x}_i\}$ from some PDF
$f(\bmath{x})$, and then moved each data point arbitrarily around its
iso-probability surface, the derived $\{\zeta_i\}$-values from the
original and `transported' set of data points would be identical,
despite the latter set clearly no longer being drawn from
$f(\bmath{x})$. Thus, the basic test embodied by the mapping
(\ref{eq:HPD_varDef}) is not consistent against all alternatives: it
is possible for there to be two rather different multidimensional PDFs
that are not distinguished by the test statistic, even in the limit of
a large number of data points. As an extreme example, consider the
two-dimensional unit circularly-symmetric Gaussian distribution
$f(x,y) = (2\pi)^{-1} \exp[-(x^2+y^2)/2]$ and the alternative
distribution $g(x,y)= \delta(y)H(x)x\exp(-x^2/2)$, where $\delta(y)$
is the Dirac delta function and $H(x)$ is the Heaviside step
function. Our basic test cannot distinguish between these two
distributions, since data points drawn from the latter can be obtained
by moving each data point drawn from the former along its
iso-probability contour to the positive $x$-axis.

There are, however, a number of ways in which this difficulty can be
overcome. Most straighforwardly, one can extend the basic approach to
include separate tests of each lower-dimensional marginal of
$f(\bmath{x})$ against the corresponding marginal data set. In a
two-dimensional case, for example, one would also test, separately,
the hypotheses that the one-dimensional data-sets $\{x_i\}$ and
$\{y_i\}$ are drawn from the corresponding reference distributions
obtained by marginalization of $f(x,y)$ over $x$ and $y$,
respectively. This would clearly very easily distinguish between the
two example distributions $f(x,y)$ and $g(x,y)$ mentioned above.

\begin{figure}
  \begin{center}
\includegraphics[height=6.6cm]{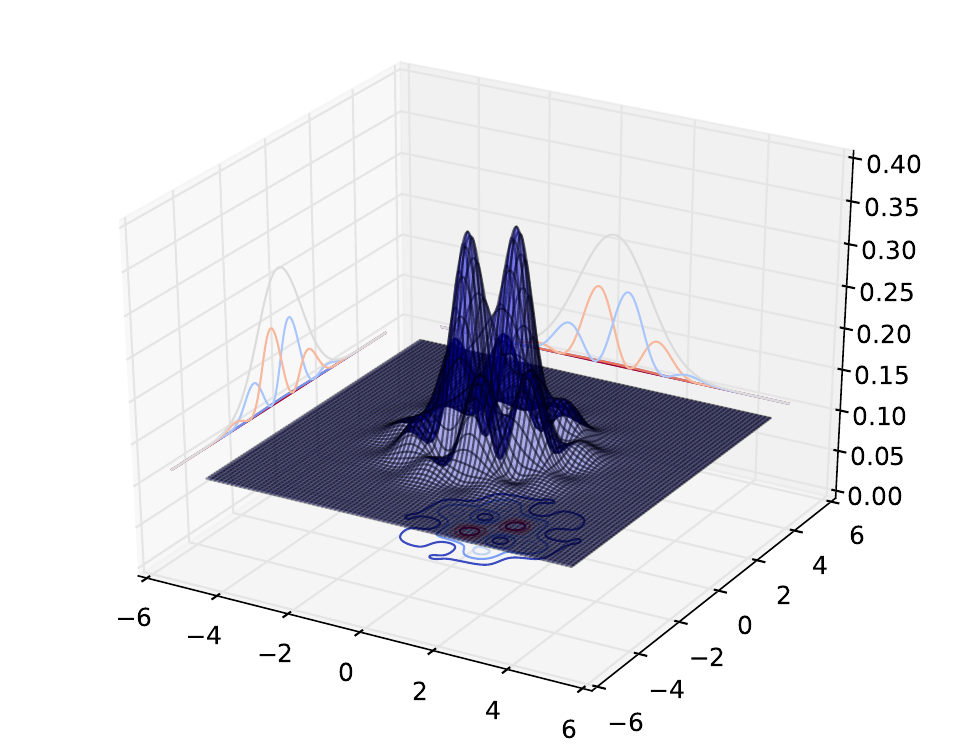}
\caption{Illustration of the `jelly mould' function
  $g(x,y) = \frac{1}{2\pi}\exp\left(-\frac{x^2+y^2}{2}\right) \left[1 + \sin(\pi
  x) \sin(\pi y)\right]$. A constant value of $0.1$ has been added to
  help visualise the function.
    \label{fig:Bastard} }
  \end{center}
\end{figure}

In practice, the above approach serves to distinguish between the
original PDF $f(\bmath{x})$ and any `reasonable' alternative PDF that
one might expect to occur in realistic scenarios. Nonetheless, one can
construct contrived alternative PDFs that cannot be distinguished from
$f(\bmath{x})$ in this way, even in the limit of a large number of
samples. An obvious approach to consructing such PDFs is take the
original PDF $f(\bmath{x})$ and redistribute probability mass around
its iso-probability surfaces in a symmetric manner that does not alter
any of its marginals. As an example, again consider the
two-dimensional unit circularly-symmetric Gaussian distribution
$f(x,y)$ discussed above, and the alternative PDF
\begin{equation}
g(x,y) = \frac{1}{2\pi}\exp\left(-\frac{x^2+y^2}{2}\right) \left[1 + \sin(\pi x) \sin(\pi y)\right],
\label{eqn:jmdef}
\end{equation}
which we call the `jelly mould' function and is plotted in
Fig.~\ref{fig:Bastard} (added to a constant value of 0.1, to help
visualise the function). Even by extending our basic approach to
include separate tests on each one-dimensional marginal, as discussed
above, one cannot distinguish between these two distributions.

To accommodate even such contrived examples, one must make use of the
fact that, if a set of data points are drawn from for some reference
distribution $f(\bmath{x})$, they will be uniformly distributed over
any given iso-probability surface of $f(\bmath{x})$.  Unfortunately,
the direct application of this result is impossible, since an
iso-probability surface is a set of zero measure.  Nonetheless, one
can make use of it approximately by artificially `discretising' the
reference distribution $f(\bmath{x})$, as follows. One first divides
the range of the distribution into $N_\mathrm{B}$ bins of width
$\Delta_f = (f_\mathrm{max}-f_\mathrm{min})/N_\mathrm{B}$, which, in
turn, allows one to identify the disjoint and exhaustive set of regions
${\cal R}_b$ $(b=1,2,\ldots,N_\mathrm{B})$ in $\bmath{x}$-space
defined by\footnote{Another possibility is to divide the range of
  $f(\bmath{x})$ into $N_{\textrm{B}}$ bins with, in general, unequal
  widths $\Delta_f^{(b)}$, which are chosen such that each
  corresponding region ${\cal R}_b$ contains the same probability
  mass. This would ensure that each region contains a similar number
  of {color{red} data points}, but could lead to regions within which the value of
  $f(\bmath{x})$ varies considerably (i.e.\ $\Delta_f^{(b)}$ is large),
 and so data points will be far
  from uniformly distributed. We will investigate this alternative
  approach in future work.}
\begin{equation}
{\cal R}_{\textrm{b}} = \{\bmath{x}:\,f_\mathrm{min}+(b-1)\Delta_f \le f(\bmath{x})
< f_\mathrm{min}+b\Delta_f\}.
\end{equation}
One then simply replaces the true reference distribution
$f(\bmath{x})$ with its 'ziggurat' approximation defined
by
\begin{equation}
f_{\textrm{Z}}(\bmath{x};N_\mathrm{B}) = \langle f(\bmath{x}) \rangle_{{\cal R}_b}\quad\mbox{if $\bmath{x} \in {\cal R}_b$}.
\label{eqn:ziggurat}
\end{equation}
In the limit $N_\mathrm{B} \to \infty$ (or $\Delta_f \to 0$), one
clearly recovers the original reference distribution
$f_{\textrm{Z}}(\bmath{x};N_\mathrm{B}) \to f(\bmath{x})$, but this is of
no use for the reason outlined above. For large, but finite, values of
$N_\mathrm{B}$, however, one obtains a good approximation to the original
reference distribution that consists entirely of a large number of
perfect `plateau' regions. One can then apply the approach described
in Section~\ref{sec:pathological} for such (deliberately constructed)
`pathological cases'. In particular, a necessary condition for the set
of values $\{\zeta_i\}$ corresponding to the data points
$\{\bmath{x}_i\}$ to be uniformly distributed is that the data points
within each perfect plateau region ${\cal R}_b$ are themselves
uniformly distributed, which will be the case (to the level of
approximation inherent to the ziggurat) if they are drawn from the
reference distribution $f(\bmath{x})$. Clearly, as one allows
$N_\mathrm{B}$ to become smaller, the ziggurat approximation
(\ref{eqn:ziggurat}) to the true reference distribution becomes
increasingly poor, and so the derived values $\{\zeta_i\}$ will, in
general, no longer be uniformly distributed. Nonetheless, there should
be a range of (large, but finite) $N_\mathrm{B}$ values, dependent both
on the form of the reference distribution $f(\bmath{x})$ and the
number of data points $\{\bmath{x}_i\}$, for which the corresponding
derived values $\{\zeta_i\}$ are (approximately) uniformly distributed
{\em only} under the null hypothesis, thus making the test sufficient
in the limit of a large number of data points.

\begin{figure}
\begin{center}
\includegraphics[height=6.8cm]{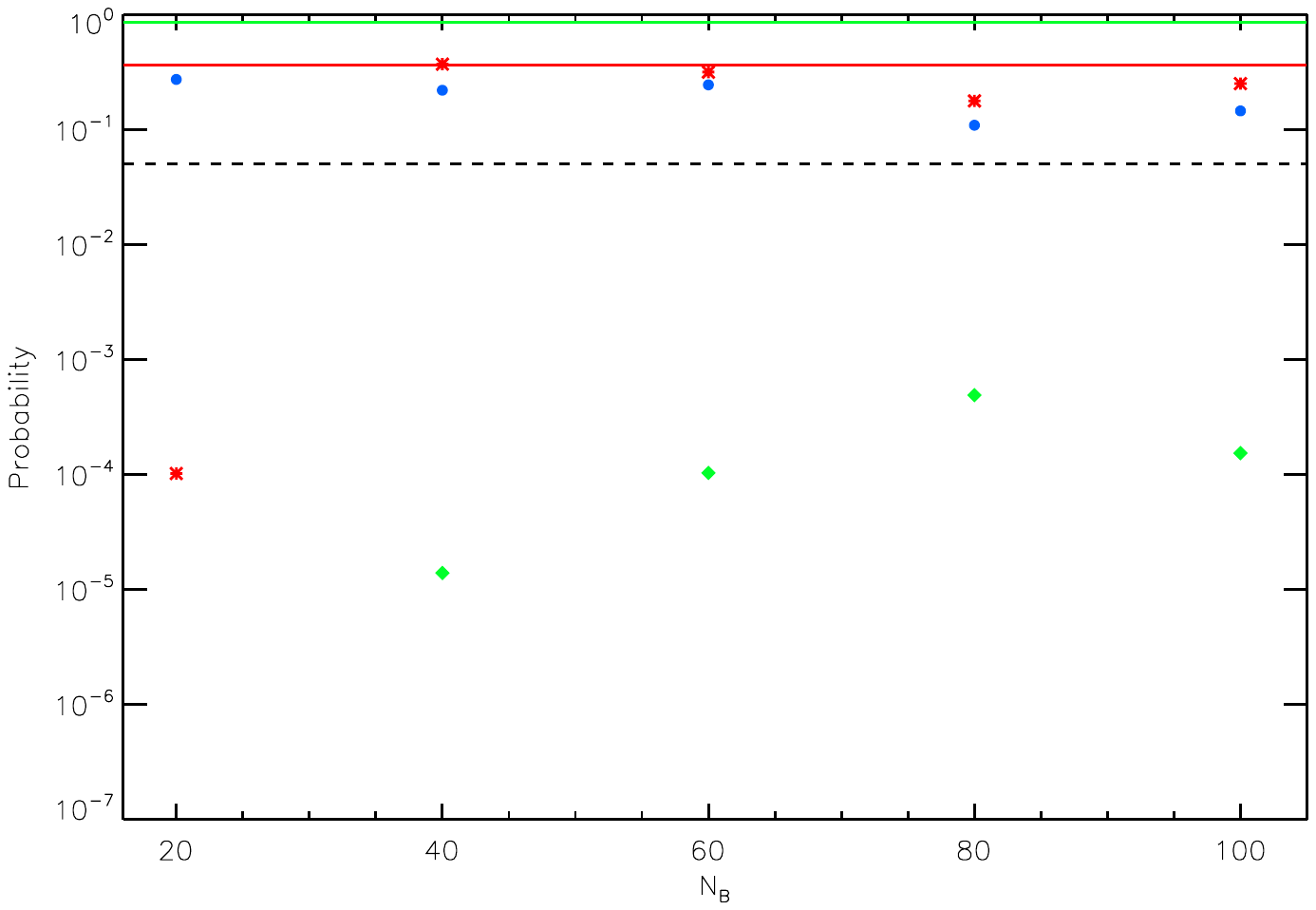}
\caption{The K-S test probability that a set of samples
  is drawn from the two-dimensional unit circularly-symmetric Gaussian
  reference distribution $f(x,y)$. The solid red and greens lines show
  the probabilities obtained using our basic test on a set of 100,000
  samples from the reference distribution $f(x,y)$ and the jelly mould
  distribution $g(x,y)$, respectively. The dashed black line shows the
  nominal 5 per cent probability level, constituting the pass/fail
  boundary of the test. The red stars show the probabilities obtained
  for the same set of samples from $f(x,y)$ in the case where the
  reference distribution is instead the 'ziggurat' approximation
  $f_Z(x,y;N_\mathrm{B})$ to the Gaussian, defined in
  (\ref{eqn:ziggurat}), for several values of $N_\mathrm{B}$.  The
  green diamonds show the corresponding probabilities obtained for the
  samples from the jelly mould $g(x,y)$.  The blue circles show the
  corresponding probabilities obtained for a set of 100,000 samples
  taken from the 'ziggurat' approximation $f_Z(x,y;N_\mathrm{B})$
  itself.}

\label{fig:csr_zigtest} 
\end{center}
\end{figure}

Figure~\ref{fig:csr_zigtest} shows a demonstration of
  this test. Two sets of data points, each containing 100,000 samples,
  were drawn: one set from the two-dimensional unit
  circularly-symmetric Gaussian distribution $f(x,y)$, and the other
  from the `jelly mould' distribution $g(x,y)$ given in
  (\ref{eqn:jmdef}). Each set of samples was tested against the
  Gaussian reference distribution $f(x,y)$. Owing to the
  iso-probability degeneracy, the basic test finds that each set of
  samples is consistent (at better than the nominal 5 per cent
  confidence level indicated by the dashed black line) with being
  drawn from the Gaussian distribution, as illustrated by the solid
  red and green lines in the figure. If, however, one instead uses the
  ziggurat approximation $f_Z(x,y;N_\mathrm{B})$ to the Gaussian as
  the reference distribution, then one can easily distinguish between
  the two sets of samples. As shown by the red stars in
  Figure~\ref{fig:csr_zigtest}, the samples from the Gaussian
  distribution pass the new test for large values of $N_\mathrm{B}$,
  but fail it for small $N_\mathrm{B}$, as the ziggurat approximation
  to the Gaussian becomes poorer. By contrast, as shown by the green
  diamonds, the samples from the jelly mould
  distribution $g(x,y)$ fail the test for all values of
  $N_\mathrm{B}$. As an additional check, the blue circles in the
  figure show the probabilities obtained for a set of 100,000 samples
  drawn from the ziggurat approximation $f_Z(x,y;N_\mathrm{B})$
  itself; since this is the reference distribution, these samples
  should pass the test for all values of $N_\mathrm{B}$, as is indeed
  the case.

It should be noted that it is not possible to extend our basic test in
the manner outlined above for the extreme case, described in
Section~\ref{sec:multipletests}, where one wishes to perform multiple
simultaneous tests but one has just a single data point drawn from
each reference distribution. This therefore precludes the application
of the extended test described above to the validation of posterior
distributions in Bayesian inference, which is discussed in
Section~\ref{sec:validpost}. In this application, one is thus
restricted to using our basic test, the passing of which provides only
a necessary, but not sufficient, condition for the Bayesian analysis
to be consistent. Indeed, for the remainder of this paper, we will
consider only our basic test.

\section{Validation of Bayesian inference analyses}
\label{sec:Validation}

\subsection{Bayesian inference}
\label{sect:BayInf}
The Bayesian system of inference is the only one that provides a
consistent extension of deductive logic $(0=\mbox{false}, 1 =
\mbox{true})$ to a broader class of `degrees-of-belief' by
mapping them into the real interval $[0,1]$ \citep{Jaynes}.
The basic rules of this extended logic allow one to write an equation
that relates the posterior probability of a set of parameters
$\vect{\theta}$ given the data $\vect{d}$ to the underlying hypothesis
$H$ that embodies the data model, and the background information $I$;
the latter comprises every assumption not explicitly stated, such as
the parametrisation of the data or the instrumental set-up used to
gather it. This equation is known as Bayes' theorem,
\begin{equation}
\Pr(\btheta | \vect{d}, H, I) =
\frac{\Pr(\vect{d}|\,\vect{\theta},H, I)\Pr(\vect{\theta}|H, I)}
{\Pr(\vect{d}|H, I)},
\label{eq:BI_BayesTheorem}
\end{equation}
where $\Pr(\vect{\theta}|\vect{d},H,I) \equiv \mathcal{P}(\btheta)$ is
the posterior distribution, $\Pr(\vect{d}|\,\mathbf{\theta},H,I)
\equiv \mathcal{L}(\btheta)$ is the likelihood,
$\Pr(\btheta|H,I)\equiv \pi(\btheta)$ is the prior, and
$\Pr(\vect{d}|H,I) \equiv \mathcal{Z}$ is a constant, known as
Bayesian evidence, which may be written as
\begin{equation}
\mathcal{Z} = \int{\mathcal{L}(\vect{\theta})\,\pi(\vect{\theta})}\,d^n\vect{\theta},
 \label{eq:BI_EvidDef}
\end{equation}
where $n$ is the dimensionality of the parameter space.

\subsection{Validation of posterior distributions}
\label{sec:validpost}

Conditioned on the assumed model $H$ (and background information $I$),
all valid Bayesian inferential statements about the parameters
$\btheta$ are encapsulated in the posterior distribution
$\mathcal{P}(\btheta)$, which combines the information provided by the
data and any other information about the parameters contained in the
prior.

Nonetheless, as discussed in the Introduction, relatively little
consideration has been given to assessing whether the derived
posterior distribution is a truthful representation of the actual
constraints on the parameters that one can infer from the data, in the
context of a given model. The possibility that some difference might
exist is the result of the almost inevitable need to use some
approximate data model $H$ in the inference process, at least in most
real-world problems. More prosaically, such differences might also be
the result of errors in the implementation.

Our proposed validation procedure for testing the assumptions made
(explicitly or implicitly) in the inference process (as well as the
software implementation) is to make use of the fact that one may
typically generate simulations of greater sophistication and realism
than the data model $H$ assumed in the analysis. One may regard the
simulations as representing some alternative, more realistic data
model $H'$.  Our approach is then to test whether the posterior
distribution(s) obtained from the simulated data by assuming $H$ are
consistent with the simulations generated assuming $H'$.

Our approach is as follows. One first generates $N_\mathrm{d}$ sets of
simulated data $\{\vect{d}^{(k)}\}$ $(k=1,2,\ldots,N_\mathrm{d})$. The
assumed true values $\btheta^{(k)}_\ast$ of the model parameters may
be the same for each simulation generated or differ between them,
depending on the nature of the inference problem. Each simulated
data-set is then analysed using the inference process under
investigation to obtain the resulting posterior distributions
$\mathcal{P}_k(\btheta)\equiv \Pr(\btheta|\vect{d}^{(k)},H,I)$. One
then applies our multiple simultaneous version of the $n$-dimensional
K-S test, as outlined in Section~\ref{sec:multipletests}, to test the
null hypothesis that each set of assumed parameter values
$\btheta^{(k)}_\ast$ is drawn from the corresponding derived posterior
$\mathcal{P}_k(\btheta)$. Thus, one calculates just a single value of
the HPD probability content for each data set, which is given by
\begin{equation}
\zeta_k \equiv \zeta(\btheta^{(k)}_\ast) = \int_{\mathcal{P}_k(\btheta) \ge
  \mathcal{P}_k(\btheta^{(k)}_\ast)} \mathcal{P}_k(\btheta)\,d^n\btheta.
\label{eqn:bayeszeta}
\end{equation}
In real-world problems, the output of the Bayesian analysis of each
data-set is usually a set of posterior-distributed samples produced by
some Markov chain Monte Carlo (MCMC) \citep[see
  e.g. ][]{Mackay_2003_information} or nested sampling
\citep{SiviaG,HobsonMCMC2007} algorithm. One also usually has access
to the value of the posterior at each sample point, in which case an
elegant proxy for (\ref{eqn:bayeszeta}) is simply the fraction of such
samples for which $\mathcal{P}_k(\theta) \ge
\mathcal{P}_k(\btheta^{(k)}_\ast)$. The set of values $\{\zeta_k\}$
$(k=1,2,\ldots,N_\mathrm{d})$ so obtained are then tested against the
standard uniform distribution using the one-dimensional K-S test.

It is worth noting that this procedure does not suffer from the usual
complications associated with the number of degrees of freedom being
reduced by the number of parameters that have been optimised. A
pertinent example of this issue is provided by Lilliefors' variant of
the K-S test for the null hypothesis that a set of data points are
drawn from a Gaussian distribution of {\em unknown} mean and variance
\citep{Lilliefors1967}. In this test, the mean and variance of the Gaussian
reference distribution used in the K-S test are first estimated from
the data points, with the result that the distribution of the K-S
distance (\ref{eqn:ksdistance}) in this case is stochastically smaller
than the usual K-S distribution (\ref{eqn:ksdistribution}). By
contrast, in applying the K-S test to the set of values $\{\zeta_k\}$
$(k=1,2,\ldots,N_\mathrm{d})$ defined in (\ref{eqn:bayeszeta}), no
optimisation of parameters has been performed. Rather, one has
calculated the full posterior distribution $\mathcal{P}_k(\btheta)$ on
the parameters for each simulated data-set. If the Bayesian inference
process is valid, then the true parameter values $\btheta^{(k)}_\ast$
should simply be drawn from precisely this posterior distribution,
which constitutes our null hypothesis.

It should be noted that our validation procedure as described here allows for the verification of the implementation and any simplifying assumptions of the data model used. It does not validate whether the simulated data model is consistent with reality. If one wishes to use this procedure for this purpose, then the reader should be aware that the number of degrees of freedom would be reduced, since then we would be using a distribution derived from the data to fit the data.

\subsection{Reparameterization invariance}
\label{sec:validrepar}

As discussed in Section~\ref{sect:BayReParam}, HPD regions are not
reparameterization invariant\footnote{The same criticism also applies
  to maximum aposteriori probability (MAP) estimators; in contrast,
  maximum-likelihood (ML) estimators are reparameterization
  invariant.}, and as a result the significance of our
multidimensional K-S test will also depend on the choice of
parameterization in the data model. In the context of posterior
distributions, however, \cite{druilhet2007} have suggested a simple,
yet general, way to overcome this problem.

The reason why HPD regions and MAP estimators do not share the
reparameterization invariance of ML estimators is the presence of the
Jacobian determinant factor in the transformation law for the prior
$\pi(\btheta)$, whereas this factor is absent from the transformation
law of the likelihood $\mathcal{L}(\btheta)$. This is a consequence of
the posterior density $\mathcal{P}(\btheta) \propto
\mathcal{L}(\btheta)\pi(\btheta)$ being defined with respect to the
Lebesgue measure on $\btheta$ \citep[see,
  e.g.,][sec. 5.7]{lehmann2006testing}. One is free, however, to
choose a different measure (or equivalently the unit of length or
volume), but this should depend (at most) only on the likelihood.
\cite{druilhet2007} suggest the use of the Jeffreys measure, which
has the density $|\mathbfss{I}(\btheta)|^{1/2}$ with respect to the
Lebesgue measure, where $\mathbfss{I}(\btheta)$ is the Fisher
information matrix. Under certain regularity conditions, the elements
of this matrix are given by
\begin{equation}
I_{ij}(\btheta) = \left\langle
\frac{\partial\ln\mathcal{L}}{\partial\theta_i}
\,\frac{\partial\ln\mathcal{L}}{\partial\theta_j}
\right\rangle_{\mathcal{L}(\btheta)}
=-\left\langle \frac{\partial^2\ln\mathcal{L}}
{\partial\theta_i\,\partial\theta_j}
\right\rangle_{\mathcal{L}(\btheta)}.
\label{eqn:fisherdef}
\end{equation}

The posterior density of $\btheta$ with respect to the Jeffreys
measure on $\btheta$, which we denote by
$\mathcal{P}_\mathrm{J_\theta}(\btheta)$, is therefore given by
\begin{equation}
 \label{eq:BI_InvIntervals}
\mathcal{P}_\mathrm{J_\theta} (\btheta) \propto \mathcal{L}(\btheta)\, |\mathbfss{I}(\btheta)|^{-1/2}
\,\pi(\btheta).
\end{equation}
Under a transformation of variables $\btheta \to
\bphi=\bphi(\btheta)$, it is straightforward to show that the Fisher
information matrix transforms as $\mathbfss{I}(\bphi) =
\mathbfss{J}^\mathrm{t} \mathbfss{I}(\btheta) \mathbfss{J}$, where
$\mathbfss{J}$ is the Jacobian matrix of the transformation, with
elements $J_{ij} = \partial \theta_i/\partial\phi_j$. Consequently,
one sees immediately that the Jacobian determinant factors in the
transformation law of the posterior (\ref{eq:BI_InvIntervals}) all
cancel, and so the posterior in the new variables is simply
\begin{equation}
\mathcal{P}_\mathrm{J_\phi}(\bphi) = \mathcal{P}_\mathrm{J_\theta}(\btheta(\bphi)).
\end{equation}
As a result, any HPD region derived from (\ref{eq:BI_InvIntervals}),
which we denote by JHPD, is invariant under smooth, monotonic
reparameterizations, such that $\mbox{JHPD}_\zeta(\bphi) =
\bphi(\mbox{JHPD}_\zeta(\btheta))$ (the corresponding MAP estimate is
also similarly invariant).  Moreover, a moment's reflection reveals
that the boundary of $\mbox{JHPD}_\zeta(\btheta)$ transforms into that
of $\mbox{JHPD}_\zeta(\bphi)$. In the context our multidimensional
K-S test, this is sufficient for the values of $\{\zeta_i\}$ derived
from the original and transformed data sets, respectively, to be
identical, and thus so too will be the significance of the test.

As pointed out by \cite{druilhet2007}, another motivation for using
the Jeffreys measure is that it is also a classical non-informative
prior, which minimizes the asymptotic expected Kullback--Leibler
distance between the prior and the posterior distributions, provided
there are no nuisance parameters \citep{bernardo1979}. Thus, if no prior knowledge is available on the
parameters $\btheta$, we may set $\pi(\btheta)$ to be uniform and
identify the last two factors in (\ref{eq:BI_InvIntervals}) as the
non-informative Jeffreys prior.

This result implies that the significance of our test is invariant to
smooth reparameterization if non-informative priors are assigned to
the parameters. If the priors assigned to any of the free parameters
are informative, however, any important reparameterizations of the
model should be considered separately.

\subsection{Nuisance parameters and marginalisation}
\label{sec:nuisance}

It is often the case that the full parameter space is divided into
parameters of interest and nuisance parameters; we will denote this by
$\btheta=(\btheta_\mathrm{i},\btheta_\mathrm{n})$. So far we have
assumed that the validation of the posterior distribution(s) is
performed over the full parameter space $\btheta$, whereas in practice
it may be preferable first to marginalise over the nuisance parameters
$\btheta_\mathrm{n}$ and then perform the validation only in the space of the
parameters of interest $\btheta_\mathrm{i}$.

It is clear that the significance of our test will, in general, differ
between validations performed in the spaces $\btheta$ and
$\btheta_\mathrm{i}$, respectively. More importantly, performing the
validation after marginalising over the nuisance parameters
$\btheta_\mathrm{n}$ is itself quite a subtle issue.  Since the
Jeffreys measure on $\btheta$ does not distinguish between parameters
of interest and nuisance parameters, if one simply marginalises
(\ref{eq:BI_InvIntervals}) over $\btheta_\mathrm{n}$, it does {\em
  not} follow that HPD regions derived from it in
$\btheta_\mathrm{i}$-space are reparameterizaton invariant, i.e.
$\mbox{JHPD}_\zeta(\bphi_\mathrm{i}) \neq
\bphi(\mbox{JHPD}_\zeta(\btheta_\mathrm{i}))$ in general\footnote{By
  contrast, as noted in Section~\ref{sect:BayReParam}, the `intrinsic
  credible interval' proposed by \cite{bernardo2005} is marginalization invariant in this sense.}.

The most conceptually straightforward way to circumvent this
difficulty \cite[see, e.g.,][]{druilhet2007,sunberger1998} is (where possible) to expand the prior
on $\btheta$ as $\pi(\btheta_\mathrm{i},\btheta_\mathrm{n}) =
\pi(\btheta_\mathrm{n}|\btheta_\mathrm{i})\pi(\btheta_\mathrm{i})$ and
define the `marginal likelihood'
$\mathcal{L}_\mathrm{m}(\btheta_\mathrm{i}) \equiv \int
\mathcal{L}(\btheta_\mathrm{i},\btheta_\mathrm{n})
\pi(\btheta_\mathrm{n}|\btheta_\mathrm{i})\,d\btheta_\mathrm{n}$. One
may then calculate the Fisher information matrix
$\mathbfss{I}_\mathrm{m}(\btheta_\mathrm{i})$ of the marginal
likelihood $\mathcal{L}_\mathrm{m}(\btheta_\mathrm{i})$ using an
analogous expression to that given in (\ref{eqn:fisherdef}), and
define the marginalised posterior density with respect to the Jeffreys
measure as
\begin{equation}
\mathcal{P}_\mathrm{J_{\theta_\mathrm{i}}} (\btheta_\mathrm{i})
\propto \mathcal{L}_\mathrm{m}(\btheta_\mathrm{i})\, |\mathbfss{I}_\mathrm{m}(\btheta_\mathrm{i})|^{-1/2}
\,\pi(\btheta_\mathrm{i}).
\label{eqn:marghpd}
\end{equation}
Using the same argument as that following (\ref{eq:BI_InvIntervals}),
any HPD region derived from (\ref{eqn:marghpd}) is invariant to
reparameterization of the parameters of interest $\btheta_\mathrm{i}$,
and thus so too is the significance our multidimensional K-S test.

\section{Applications}
\label{sect:Applications}

We now illustrate the use of our multidimensional K-S test and its
application to the validation of Bayesian inference analyses.  In
Section~\ref{sect:ToyModel}, we apply the multidimensional K-S test
to the simple toy example of testing whether a set of two-dimensional
data $\{x_i,y_i\}$ are drawn from a given two-dimensional Gaussian
reference distribution. In Section~\ref{sect:PlanckSZcase} we apply
our related approach for validating posterior distributions to the
real-world case of estimating the parameters of galaxy clusters
through their Sunyaev--Zel'dovich (SZ) effect in the recently-released
\planck nominal mission microwave temperature maps of the sky
\citep{nominal_planck_dr}.

\subsection{Multidimensional K-S test: toy problem}
\label{sect:ToyModel}

\begin{figure*}
  \begin{center}
\includegraphics[height=4.56cm]{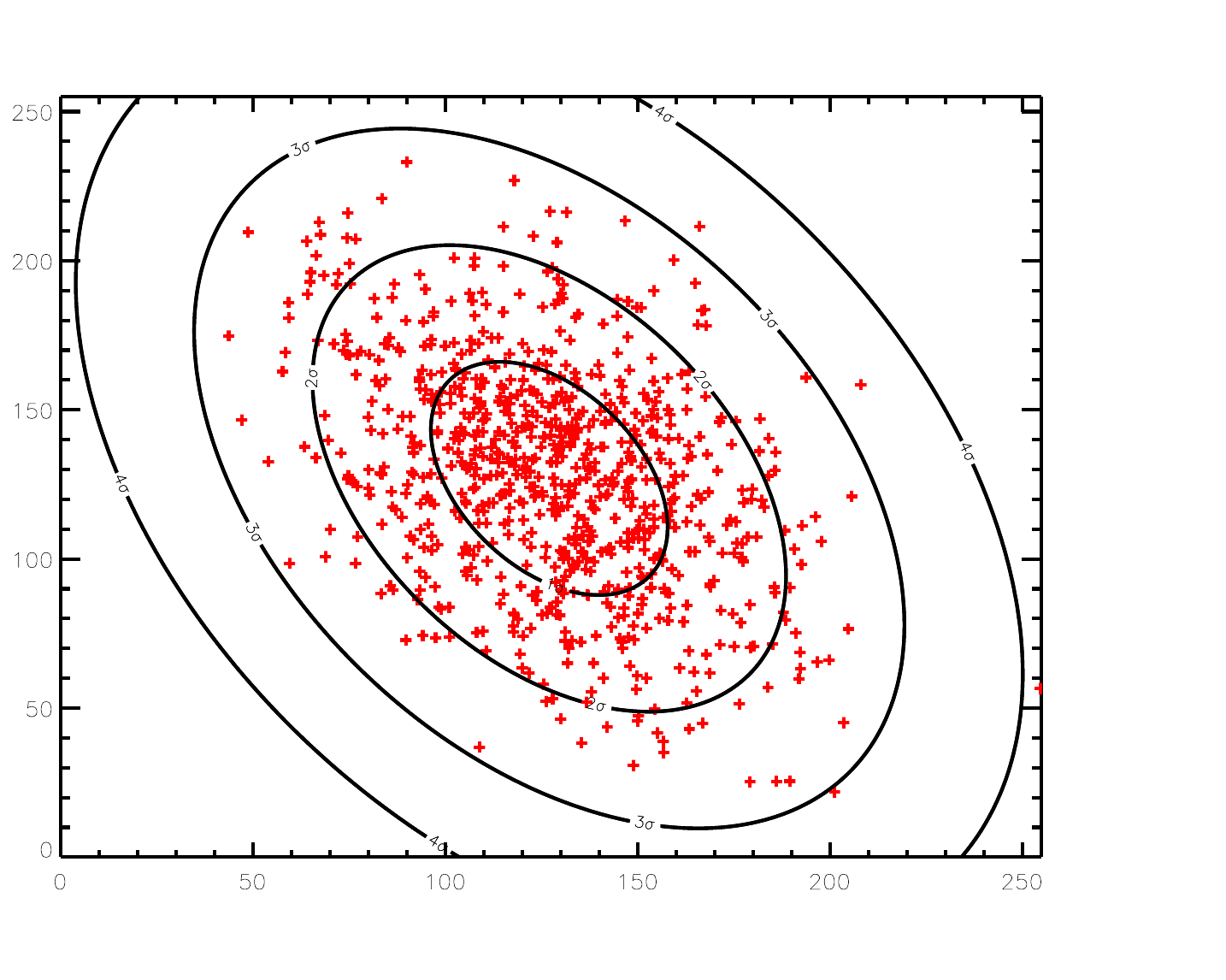}
\hspace*{-0.45cm}\includegraphics[height=4.4cm]{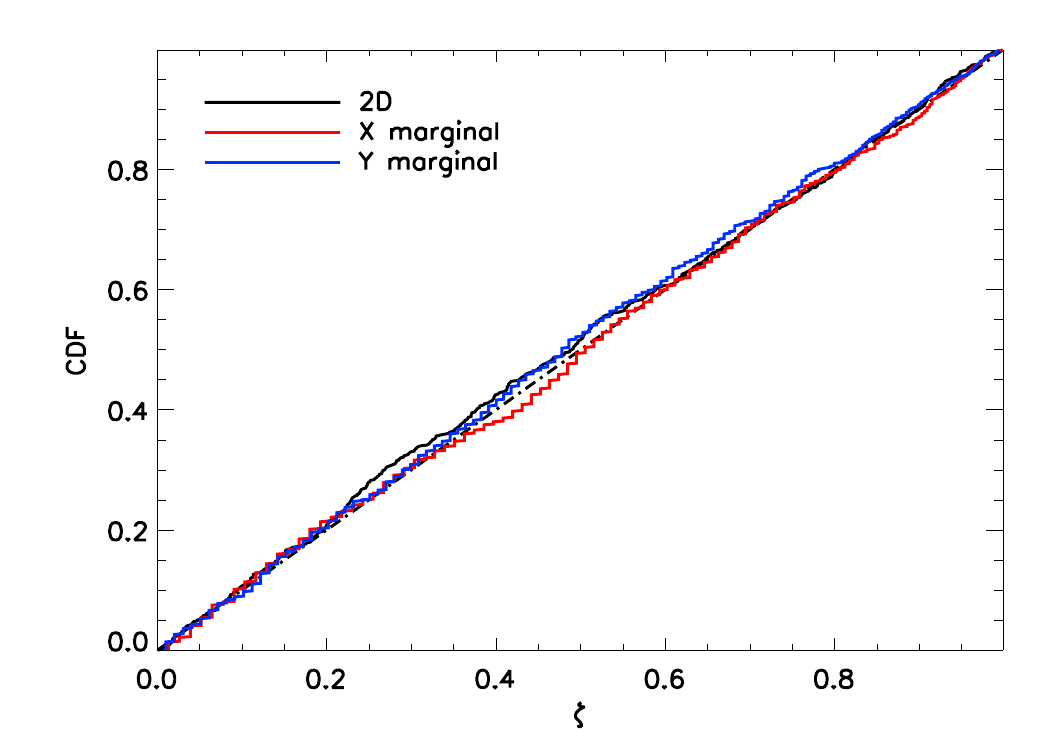}
\includegraphics[height=4.4cm]{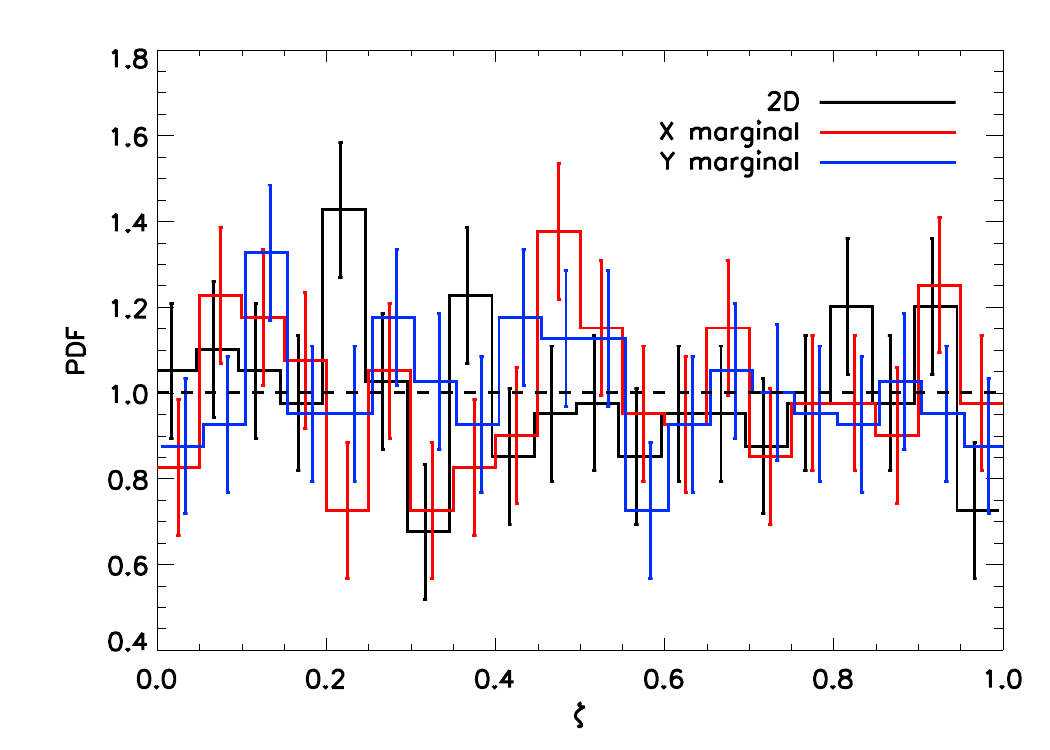}
    \caption{Left: the data points $\{x_i,y_i\}$ (red crosses) and
      2-dimensional Gaussian reference distribution $f(x,y)$ (black 1
      to $4\sigma$ contours) to which the multidimensional K-S test
      is applied. In this case the data are drawn from the reference
      distribution, so that the `correlation angle'
      $\psi=30^\circ$ and `mismatch angle'
      $\psi_\mathrm{mis}=0^\circ$. Middle: the CDF of the standard
      uniform distribution (dot-dashed black line) and the empirical
      CDFs of the corresponding $\{\zeta_i\}$-values for the full
      2-dimensional case (solid black line) and for the
      one-dimensional $x$- and $y$-marginals, respectively (red and
      blue solid lines). Right: the PDF of the standard uniform
      distribution (dashed black line) and the empirical PDFs of the
      corresponding $\{\zeta_i\}$-values (constructing by dividing
      them into 20 bins) for the full 2-dimensional case (solid black
      line) and for the one-dimensional $x$- and $y$-marginals,
      respectively (red and blue solid lines); the error bars denote
      the Poissonian uncertainty. }
    \label{fig:ToyNominal}
  \end{center}
\end{figure*}

\begin{figure*}
  \begin{center}
\includegraphics[height=4.56cm]{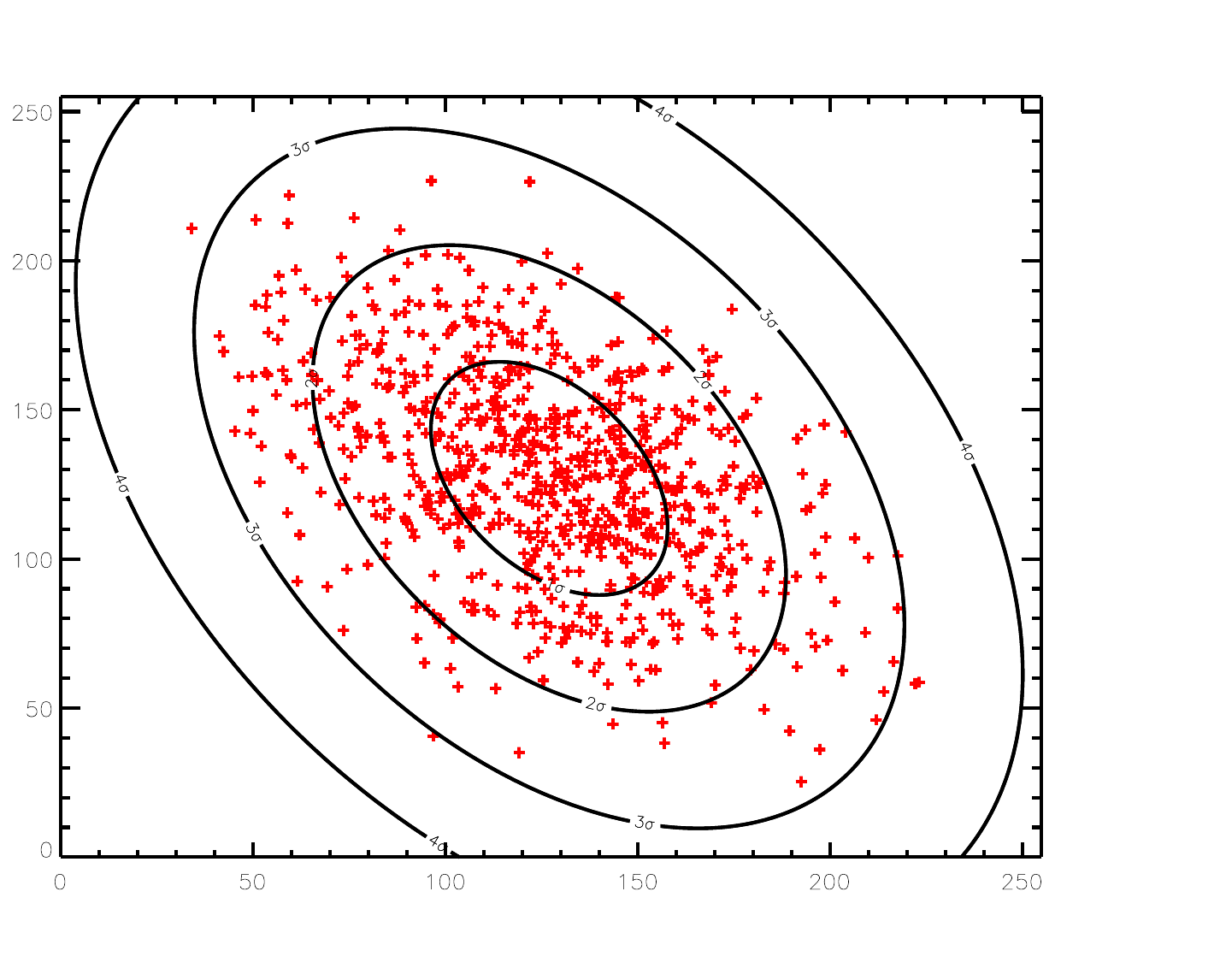}
\hspace*{-0.45cm}\includegraphics[height=4.4cm]{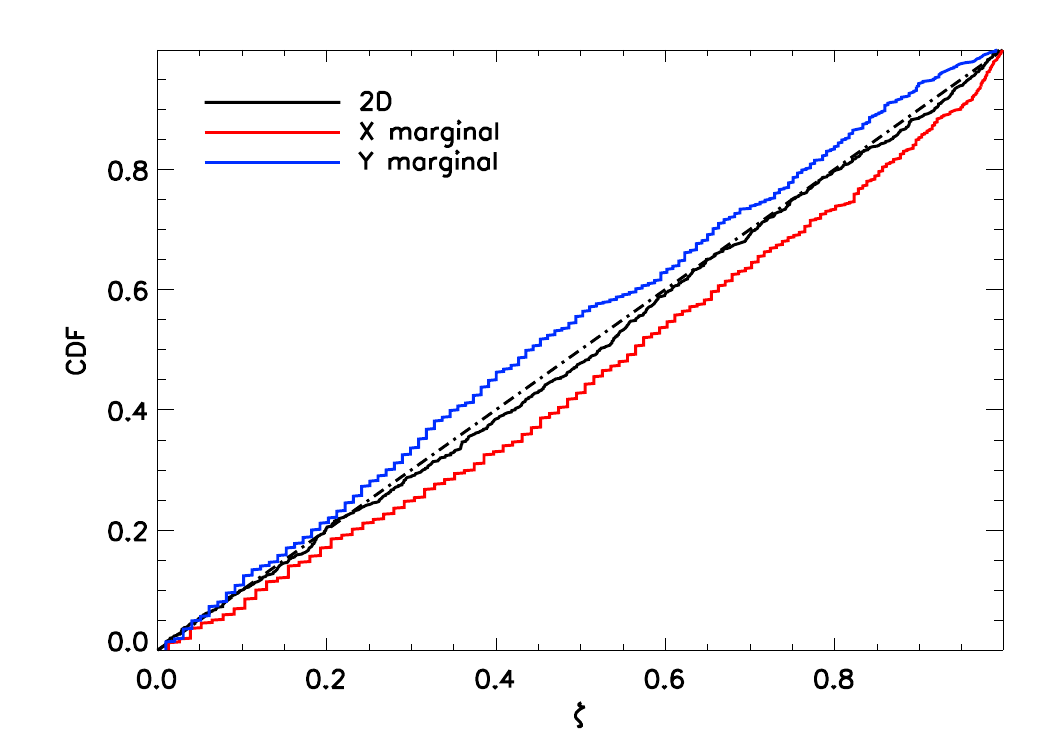}
\includegraphics[height=4.4cm]{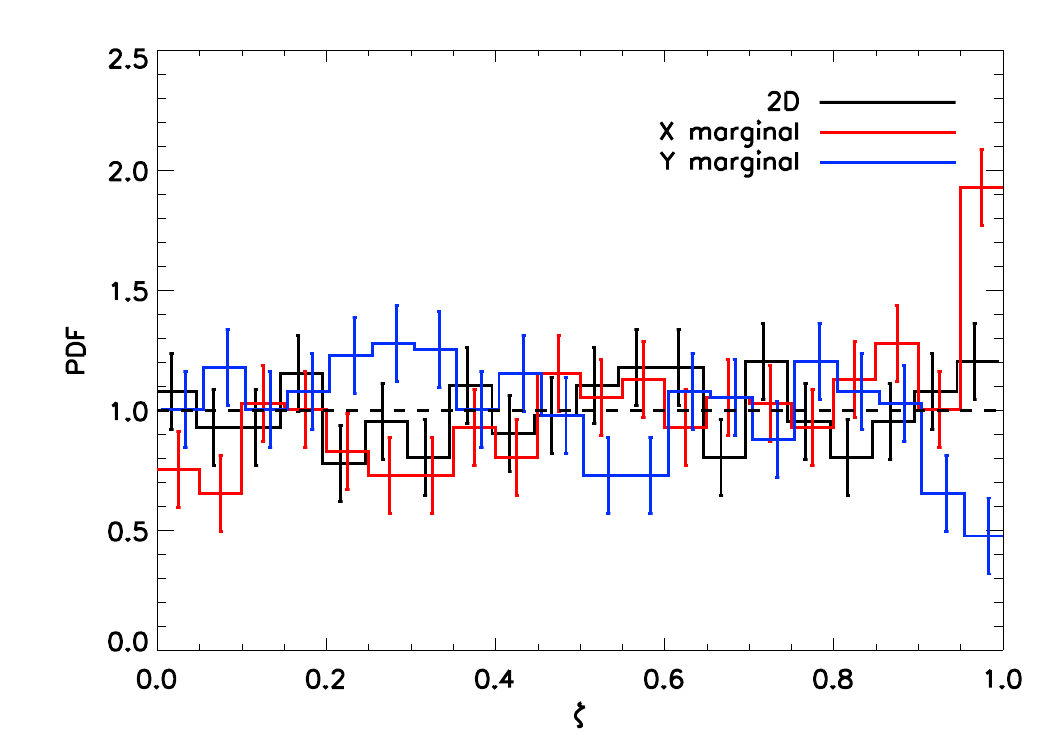}
     \caption{As in Fig.~\ref{fig:ToyNominal}, but for
       $\psi=45^\circ$ and
       $\psi_\mathrm{mis}=15^\circ$.
    \label{fig:ToyWrong15}}
  \end{center}
\end{figure*}

\begin{figure*}
  \begin{center}
\includegraphics[height=4.56cm]{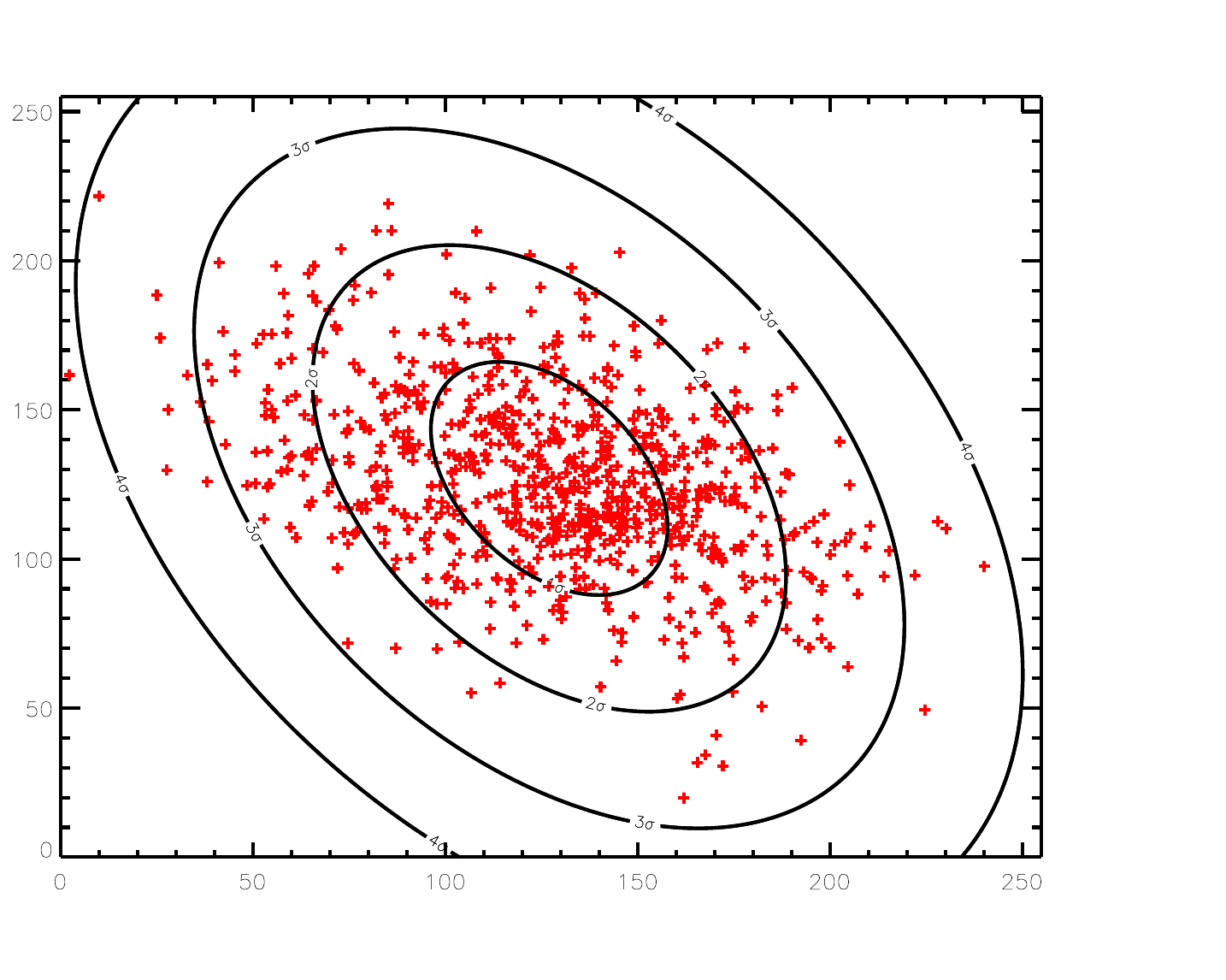}
\hspace*{-0.45cm}\includegraphics[height=4.4cm]{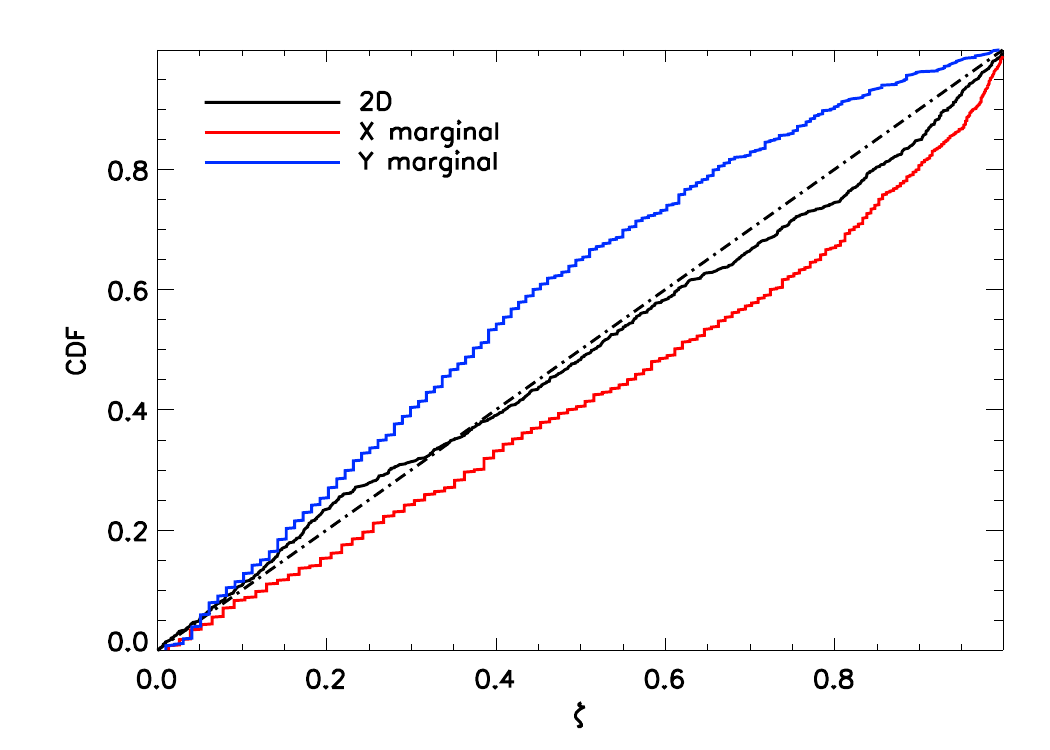}
\includegraphics[height=4.4cm]{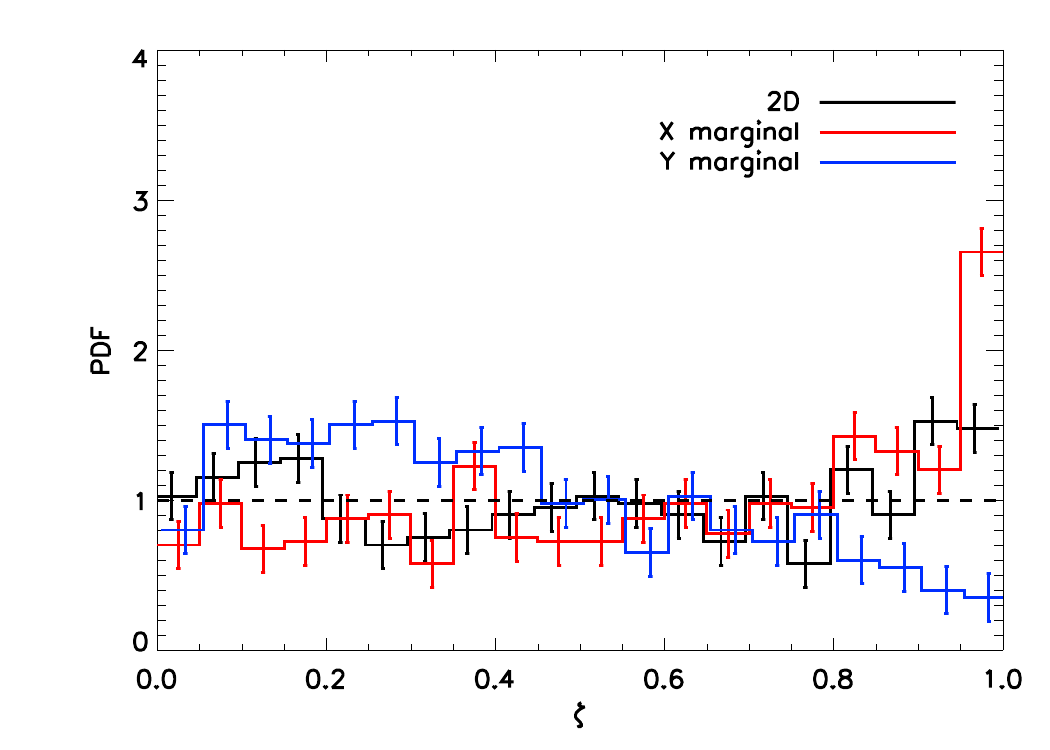}
    \caption{As in Fig.~\ref{fig:ToyNominal}, but for
       $\psi=60^\circ$ and
       $\psi_\mathrm{mis}=30^\circ$.
    \label{fig:ToyWrong30}}
  \end{center}
\end{figure*}

\begin{figure*}
  \begin{center}
\includegraphics[height=4.56cm]{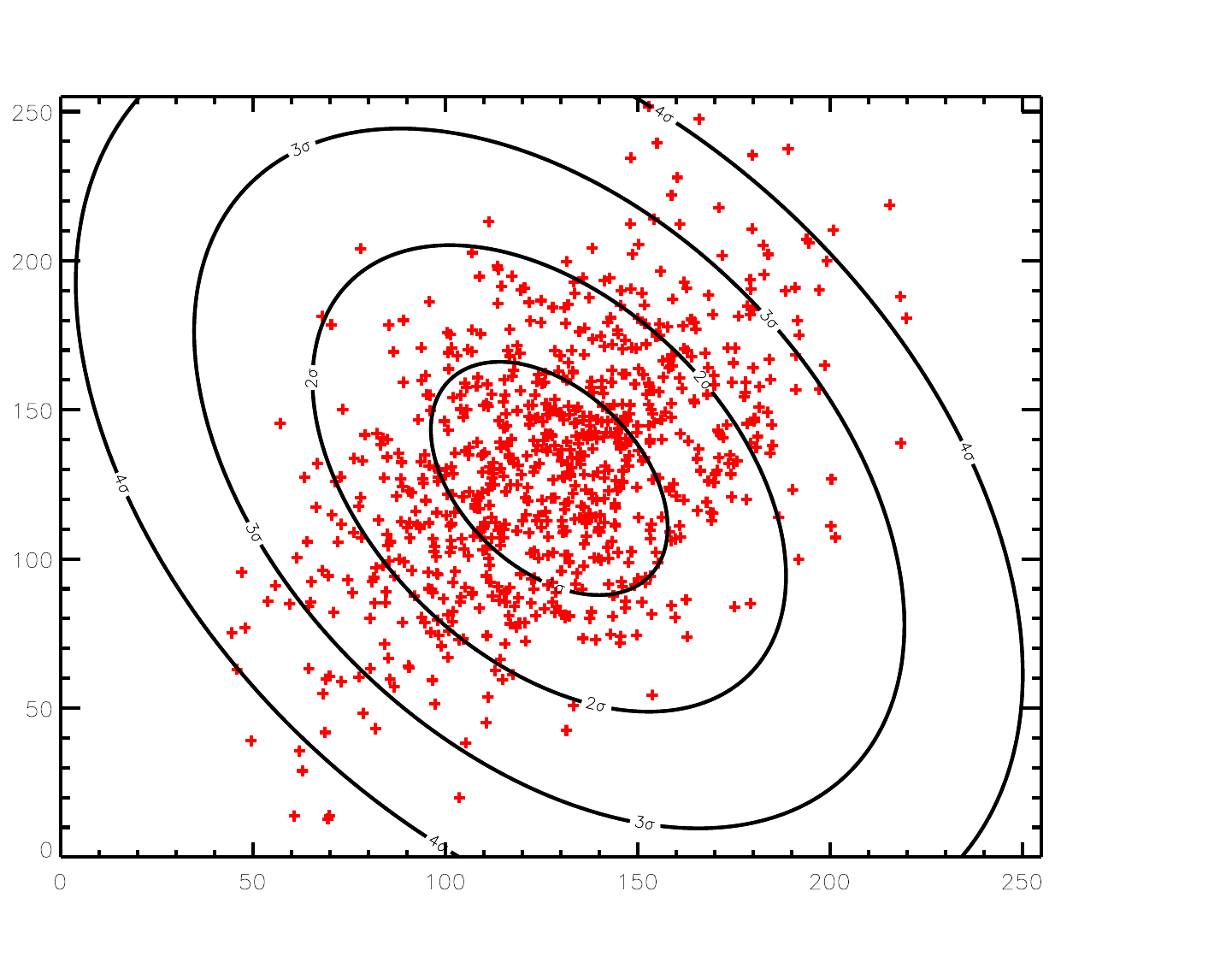}
\hspace*{-0.45cm}\includegraphics[height=4.4cm]{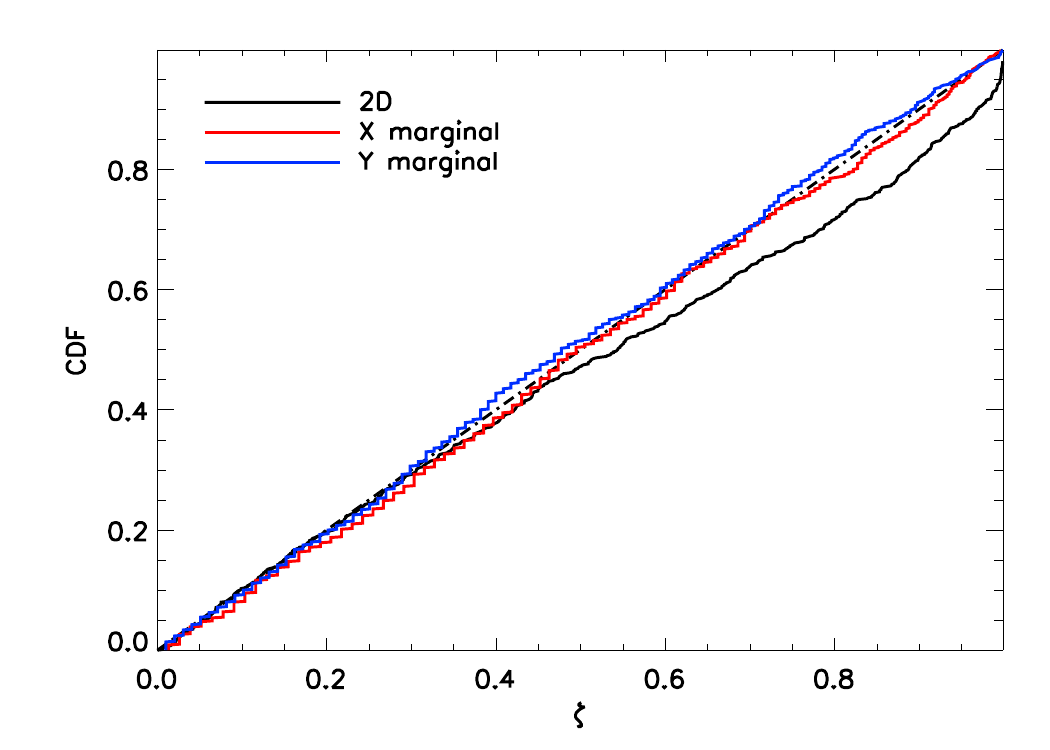}
\includegraphics[height=4.4cm]{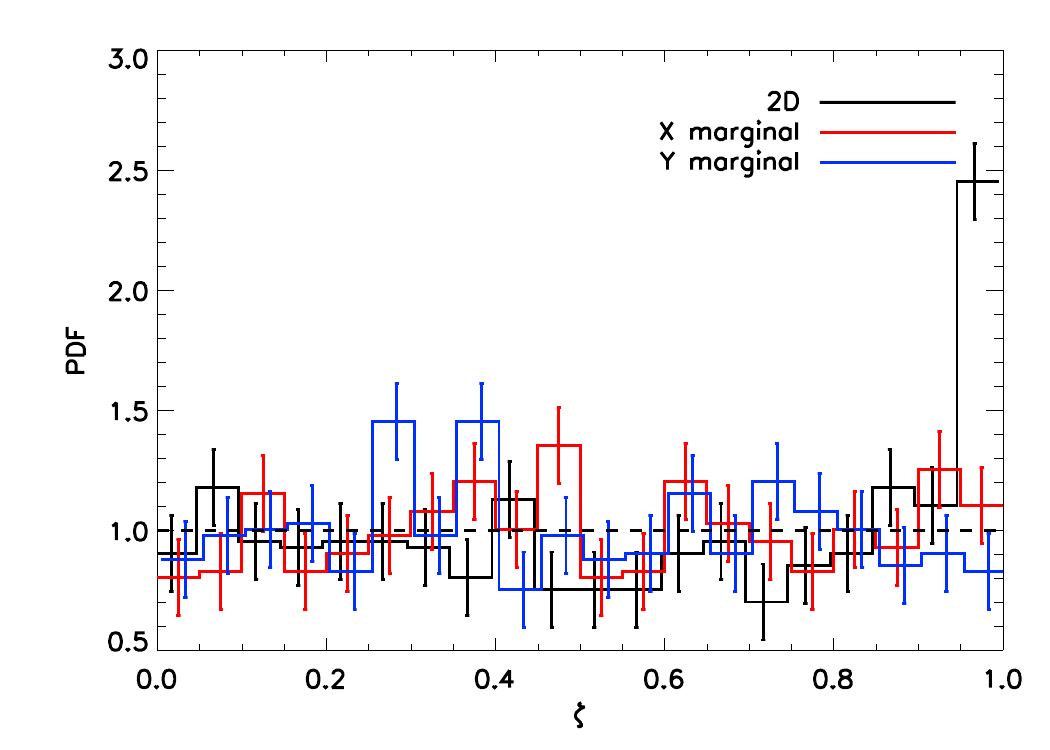}
    \caption{As in Fig.~\ref{fig:ToyNominal}, but for
      $\psi=-30^\circ$ and
      $\psi_\mathrm{mis}=-60^\circ$, so that the 2-dimensional
      Gaussian from which the data are drawn is the reflection in the
      $y$-axis of the reference distribution; hence the
      one-dimensional marginals of the two distributions are identical.
    \label{fig:ToyWrongFlipped}}
  \end{center}
\end{figure*}

We illustrate our multidimensional K-S test by first applying it,
separately, to four sets of 800 data points $\{x_i,y_i\}$ (denoted by the red crosses in the left
panels of Figs~\ref{fig:ToyNominal} to \ref{fig:ToyWrongFlipped},
respectively) to test the hypothesis that each set is drawn from the
2-dimensional Gaussian reference distribution $f(x,y)$ indicated by
the black 1 to $4\sigma$ contours in each figure. Although the first
data set is indeed drawn from $f(x,y)$, the remaining three datasets
are drawn from 2-dimensional Gaussian distributions that are rotated
in the $xy$-plane by different angles relative to $f(x,y)$.  We use
the term `correlation angle' $\psi$ to denote the angle measured
anticlockwise from the $y$-axis of the first principal direction of
the distribution from which each data set is drawn, whereas the
`mismatch angle' $\psi_\mathrm{mis}=\psi-\psi_\mathrm{ref}$, where
$\psi_\mathrm{ref}=30^\circ$ is the correlation angle of the reference
distribution $f(x,y)$; these values are listed in
Table~\ref{tab:Pvalues}.

As mentioned in Section~\ref{sec:nuisance}, the significance of our
test will, in general, be different after marginalization over some
parameters. Thus, in each of the four cases, it is of interest also to
test, separately, the hypotheses that the one-dimensional data-sets
$\{x_i\}$ and $\{y_i\}$ are drawn from the corresponding Gaussian
reference distributions obtained after marginalization of $f(x,y)$
over $x$ and $y$, respectively.

Since the marginal distributions are one-dimensional, it is also
possible to compare directly the CDFs of the data-sets $\{x_i\}$ and
$\{y_i\}$ with those of the corresponding one-dimensional Gaussian
reference marginal in a standard K-S test. We therefore also perform
these tests in each of the four cases, in order to compare the
resulting $p$-values with those obtained using our method.


\subsubsection{Mismatch angle: 0 degrees}

Figure~\ref{fig:ToyNominal} shows our nominal toy test example in
which the data points are indeed drawn from the reference
distribution.  The middle panel of Figure~\ref{fig:ToyNominal} shows
the CDF of the corresponding $\{\zeta_i\}$ values obtained from the
joint two-dimensional test (solid black line), and from the separate
tests on the $x$- and $y$-marginals (red and blue solid lines). In all
cases, the empirical CDF lies very close to that of the standard
uniform distribution (dot-dashed black line), as one would expect.
The resulting $p$-values derived from these tests are given in the
first column of Table~\ref{tab:Pvalues}, and confirm that the data
are consistent with being drawn from the reference distribution.  We
also list in parentheses the $p$-values obtained by applying the
standard one-dimensional K-S test, separately, to the $x$ and
$y$-marginals, which again are consistent with the null hypothesis.

For the purposes of illustration, in the right panel of
Fig.~\ref{fig:ToyNominal}, we also plot the empirical PDFs of the
$\{\zeta_i\}$-values in each case, constructed by dividing the range
$[0,1]$ into 20 equal-width bins; the Poissonian error-bars on each
bin are also shown. As expected, these PDFs all appear consistent with
the standard uniform distribution.  Although this plot plays no part
in deriving the significance of the test, it allows for a visual
inspection of the PDFs. As we will see later, in the case of
deviations from uniformity, this can provide useful clues as to the
nature of the discrepancy.

\begin{table}
\caption{The $p$-values for the null hypothesis, as obtained from our
  multidimensional K-S test applied to the data points (red crosses)
  and reference distribution (black contours) shown in the left panels
  of Figs~\ref{fig:ToyNominal}~to~\ref{fig:ToyWrongFlipped}.  In each
  case, the `correlation angle' $\psi$ is the angle, measured
  anticlockwise from the $y$-axis, of the first principal axis of the
  distribution from which the samples were actually drawn. The
  `mismatch angle' $\psi_\mathrm{mis}=\psi-\psi_\mathrm{ref}$, where
  $\psi_\mathrm{ref}=30^\circ$ is the correlation angle of the assumed
  reference distribution. The $p$-values are given both for the full
  two-dimensional case, and for separate tests of the corresponding
  one-dimensional marginals in $x$ and $y$, respectively. Also given,
  in parentheses, are the $p$-values obtained from the standard
  one-dimensional K-S test applied directly to the $x$- and
  $y$-marginals, respectively.}
\begin{center}
\begin{tabular}{|l|c|c|c|c|}\hline\hline
$\psi$ & $30^\circ$ & $45^{\circ}$ & $60^{\circ}$ & $-30^\circ$\\
$\psi_\mathrm{mis}$ &$0^\circ$ & $15^{\circ}$ & $30^{\circ}$ & $-60^\circ$\\
 \hline
2-D & 0.27 & 0.59 & $6.6\!\times\! 10^{-3}$& $1.4 \!\times\!
10^{-6}$ \\ $x$ marginal & 0.37 & $2.7\!\times\! 10^{-5}$ & $4.1
\!\times\! 10^{-13}$ & 0.41\\ & (0.55) & $(2.7\!\times\! 10^{-3})$ &
$(6.8 \!\times\! 10^{-11})$ & (0.93)\\ $y$ marginal & 0.41 & $1.7\!
\times\! 10^{-3}$ & $ 4.0\! \times\! 10^{-18}$& 0.51 \\ & (0.10) &
$(1.4\! \times\! 10^{-4})$ & $(3.9\! \times\! 10^{-20})$& (0.11)
\\ \hline\hline
\end{tabular}
\end{center}
\label{tab:Pvalues}
\end{table}

\subsubsection{Mismatch angle: 15 degrees}

We now consider the case where the 2-dimensional Gaussian from which
the data are actually drawn is rotated through an angle
$\psi_\mathrm{mis}=15^\circ$ relative to the reference distribution,
as shown in the left panel of Fig.~\ref{fig:ToyWrong15}. The CDFs of
the resulting $\{\zeta_i\}$-values are shown in the middle panel for
the 2-dimensional case and the one-dimensional $x$- and $y$-marginals,
respectively. In the 2-dimensional case, the empirical CDF lies close
to that of a uniform distribution, and the corresponding $p$-value
obtained is $p=0.59$ (see Table~\ref{tab:Pvalues}), which shows that
the test is not sufficiently sensitive to rule out the null
hypothesis. For the one-dimensional marginals, however, the CDFs do
appear to differ significantly from that of the uniform distribution,
and this is verified by the resulting $p$-values of $p=2.7\times
10^{-5}$ and $p=1.7\times 10^{-3}$, respectively.  This shows the
merit of performing the test both on the full joint distribution and
on the marginal distributions. For comparison, the $p$-values obtained
by applying the standard K-S test directly to the one-dimensional $x$-
and $y$-marginals, respectively, are $p=2.7 \times 10^{-3}$ and $p=1.4
\times 10^{-4}$.

The empirical binned PDFs of the $\{\zeta_i\}$-values plotted in the
right panel of Fig.~\ref{fig:ToyWrong15} are easier to interpret that
their CDFs, and provide more clues as to the nature of the
discrepancy between the data and the reference distribution.  In
particular, the large excess in the final bin for the $x$~marginal and
the corresponding decrement for the $y$~marginal are due to the
non-zero `mismatch angle' between the true and reference
distributions; this results in the marginal distributions of the data
in being too narrow in $x$ and too broad in $y$, as compared to the
marginals of the reference distribution.

\subsubsection{Mismatch angle: 30 degrees}

Fig.~\ref{fig:ToyWrong30} shows the results for the case in which the
mismatch angle is increased to $\psi_\mathrm{mis}=30^\circ$. The CDFs
plotted in the middle panel are now all visibly discrepant from that
of a uniform distribution. From Table~\ref{tab:Pvalues}, we see that
this observation is confirmed by the very small $p$-values obtained in
this case. Similarly tiny $p$-values, given in parentheses in the
table, are also obtained when applying the standard one-dimensional
K-S test directly to the $x$- and $y$-marginals. The empirical binned
PDFs plotted in the right panel exhibit similar behaviour to those
obtained for $\psi_\mathrm{mis}=15^\circ$, but are more exaggerated.
In particular, for the $x$-marginal, relative to the uniform
distribution there is now clear decrement in the PDF around $\zeta
\sim 0.2$ and an excess near $\zeta \sim 1$, with the $y$-marginal
exhibiting complementary behaviour. The behaviour of the $x$-marginal
($y$-marginal) clearly follows what one would if the spread of the
data is greater (less) than the width of the reference distribution;
if we had instead sampled the data from a distribution that was wider
(narrower) than the reference distribution (i.e. by multiplying both
eigenvalues of the reference covariance matrix by the same factor),
rather than rotated with respect to it, then {\em all} the PDFs would
have exhibited this pattern.

\subsubsection{Mismatch angle: $-$60 degrees}

Fig.~\ref{fig:ToyWrongFlipped} shows the results for the case in which
the distribution from which the data are sampled is the reflection in
the $y$-axis of the reference distribution, resulting in
$\psi_\mathrm{mis}=-60^\circ$. In this case the $x$-and $y$-marginals
of the two distributions are identical. As expected, the CDFs for the
two marginals, plotted in the middle panel, appear completely
consistent with that of a uniform distribution, which is in agreement
with the large $p$-values obtained in this case (see
Table~\ref{tab:Pvalues}). For the 2-dimensional case, however, the CDF
is clearly discrepant, and the resulting $p$-value rules out the null
hypothesis at extremely high significance. This shows, once again,
the merit of testing for consistency both of the full joint
distribution and its marginals. The PDF for the 2-dimensional case,
shown in the right panel, again shows a clear excess near $\zeta \sim
1$, relative to the uniform distribution.

\subsubsection{Comparison with theoretical expectation}

\begin{figure}
  \begin{center}
    \leavevmode
     \hspace*{-0.2cm}\includegraphics[width=0.5\textwidth]{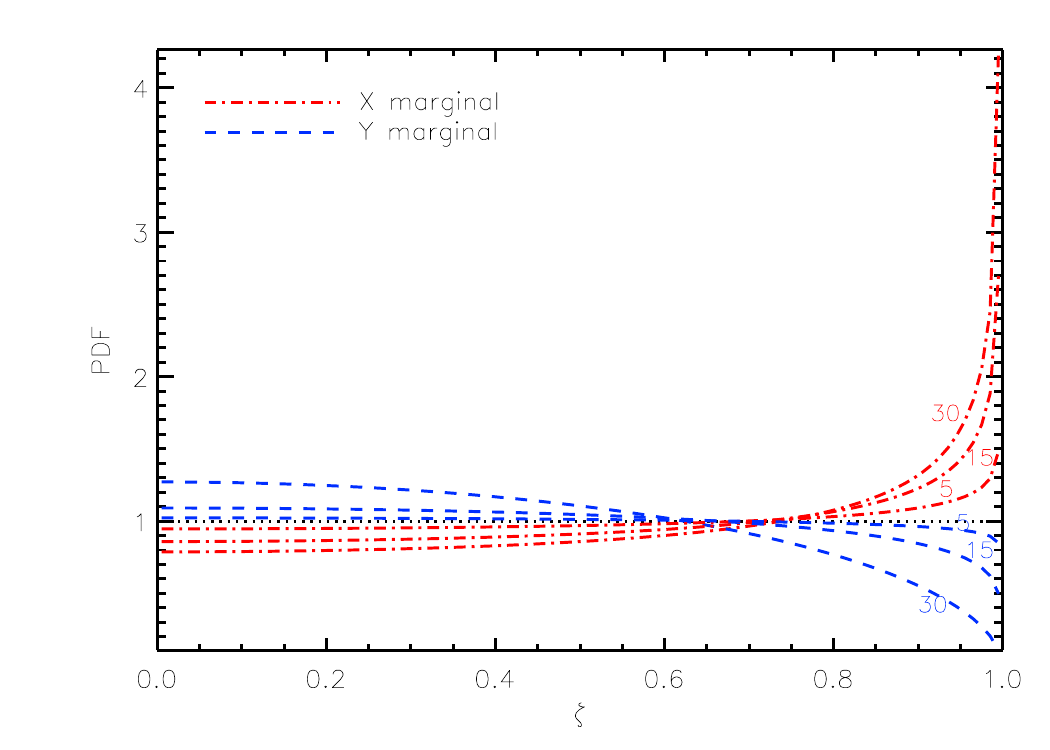}
     \caption{The expected PDF of the $\zeta$-variable for the
       $x$-marginal (dot-dashed line) and $y$-marginal (blue dashed
       line) for mismatch angles of $5^\circ$, $15^\circ$ and
       $30^\circ$.}
    \label{fig:ToyPredictions}
  \end{center}
\end{figure}

In this simple toy example, one may in fact calculate
straightforwardly the expected PDF of the $\zeta$-variable for any
given mismatch angle $\psi_\mathrm{mis}$ between the reference
distribution and the true one from which the data are drawn. Each
value of $\zeta$ specifies an HPD of the reference distribution that
itself defines a region of the parameter space. If the probability
content of the true distribution within this region is $\chi(\zeta)$, then
the PDF is simply $p(\zeta)=\chi(\zeta)/\zeta$.
Fig.~\ref{fig:ToyPredictions} shows the expected PDFs for the $x$-
and $y$-marginals for 3 different mismatch angles $\psi_\mathrm{mis}$,
which clearly exhibit the behaviour we observed previously in
Figs~\ref{fig:ToyWrong15} and~\ref{fig:ToyWrong30}.

We note that the sensitivity of our test in this toy example depends
upon the maximum distance between the CDF of the $\zeta$-values and
that of a uniform distribution, and also on the number of data
samples. It is clear that, for larger mismatch angles, fewer samples
will be necessary to reject the null hypothesis at some required level
of significance.
Indeed, from Fig.~\ref{fig:ToyPredictions}, one may estimate the number of samples $N$ required to obtain a given significance of
when testing either the $x$- or $y$-marginal.
For example, to reject the null hypothesis at a significance of 95 per cent, in the case of $\psi_\mathrm{mis}=30^\circ$, one finds
that $N \sim 130$ for both the $x$-~and~$y$-marginals, whereas for $\psi_\mathrm{mis}=5^\circ$, one requires $N\sim 2500$ for the $x$-marginal and $N\sim 17,000$ for the $y$-marginal.

\subsubsection{Testing in many dimensions}

\begin{figure}
  \begin{center}
    \leavevmode
     \hspace*{-0.2cm}\includegraphics[width=0.5\textwidth]{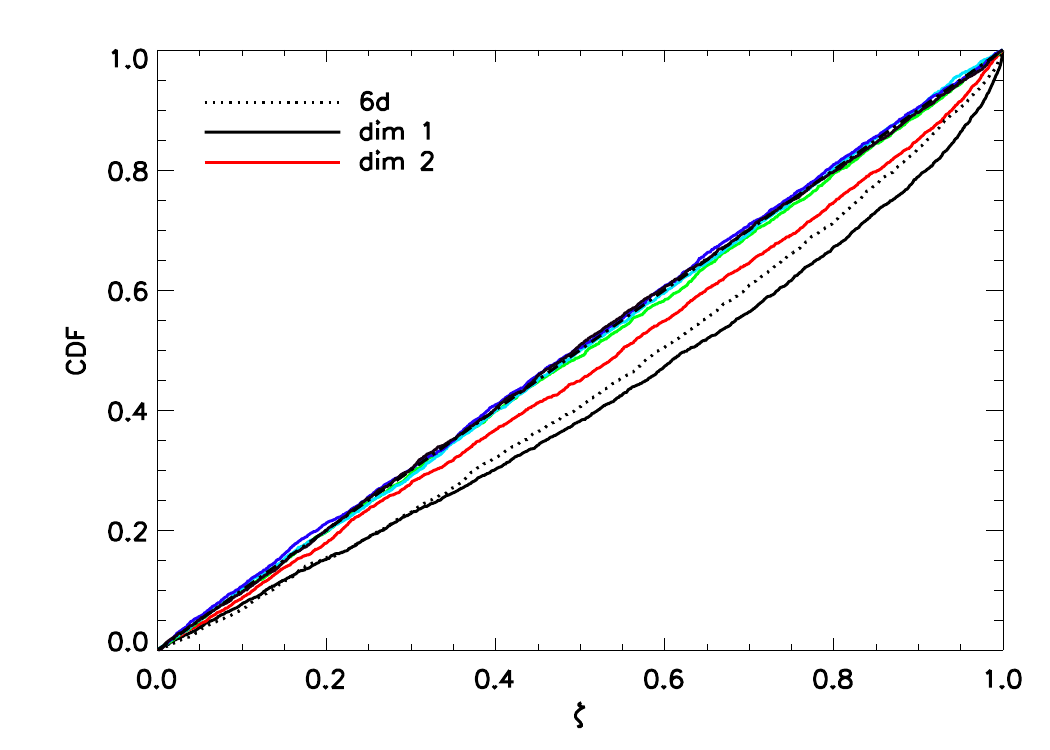}
     \caption{The CDFs of the $\{\zeta_i\}$-values for an example in 6 dimensions.  The full 6-D case is shown
     by the dotted line, the marginal for dimension 1 (with a scale parameter underestimation) is shown by the black solid line,
      and the marginal for dimension 2 (with a mean translation) is shown by the red line.  Each of these distributions
      is inconsistent with the uniform CDF, possessing a $p$-value $<10^{-8}$.  The remaining dimensions
      are consistent with the uniform CDF and possess $p$-values in the range $0.2 - 0.94$.}
    \label{fig:6d_example}
  \end{center}
\end{figure}

So far we have demonstrated our multidimensional K-S test as applied to a toy-model in two dimensions.
Application to distributions with an arbitrary number of dimensions is straightforward and we demonstrate this here by
extending the toy example to 6 dimensions.  In this example, our 6-dimensional Gaussian fiducial
distribution has two discrepancies compared to the true distribution from which the data are drawn: in the first dimension the
fiducial Gaussian $\sigma$ is underestimated relative to the true $\sigma$ by 30\%; in the second dimension, the mean of the
fiducial distribution is translated by $0.5\sigma$.
The CDFs of the 1-D marginals and the full 6-D distribution, together with their $p$-values, are shown in Fig.
\ref{fig:6d_example}.  The marginals of dimensions 1 and 2 show a strong discrepancy, with $p$-values that
robustly reject the null-hypothesis, as does the CDF of the full 6-D distribution.  The other dimensions are
consistent with no discrepancy.

\subsubsection{Simultaneous test of multiple reference distributions}

Returning to our two dimensional toy example we now demonstrate the flexibility of our approach by performing a
simultaneous test of multiple reference distributions, as discussed in
Section~\ref{sec:multipletests}. To this end, we generate simulated
data as follows. For each data point, we draw a correlation angle
$\psi$ from a uniform distribution between $30^\circ$
and $60^\circ$; this correlation angle is then used to define a
reference distribution from which a single sample is drawn.  The
process is repeated 2400 times.  This therefore corresponds to the
extreme example, discussed in Section~\ref{sec:multipletests}, in
which just a single data point is drawn from each reference
distribution.

Performing the simultaneous test on the full set of 2400 data points,
as outlined in Section~\ref{sec:multipletests}, produces the
$p$-values listed in the first column of
Table~\ref{tab:Pvalues_multi_correct}. As expected, these values
clearly support the null hypothesis that each data point was drawn
from its respective reference distribution. One may also
straightforwardly perform the test on subsets of the data, with the
division based on some given property of the corresponding reference
distributions. To this end, Table~\ref{tab:Pvalues_multi_correct} also
shows the $p$-values for three subsets of the data for which the
correlation angle $\psi$ lies in the ranges indicated. Once again, all
the $p$-values support the null hypothesis, as expected.

\begin{table}
\caption{The $p$-values for the null hypothesis, as obtained from our
  multidimensional K-S test applied simultaneously to 2400
  data points, where each one is drawn from a separate
  two-dimensional Gaussian reference distribution with a correlation
  angle $\psi$ itself drawn from a uniform distribution between
  $30^\circ$ and $60^\circ$. The $p$-values are given both for the
  full two-dimensional case, and for the $x$- and $y$-marginals,
  respectively. The left column lists the $p$-values for the full set
  of data points, and the remaining three columns list the $p$-values
  for subsets of the data for which $\psi$ lies in the
  ranges indicated.}
\begin{center}
\begin{tabular}{|l|c|c|c|c|}
\hline\hline
& $30^\circ \le \psi \le 60^\circ$ & $\psi < 40^\circ$ & $40^\circ \le \psi < 50^\circ$ & $50^\circ \le \psi$ \\
 \hline
 2-D & 0.51 & 0.41 & 0.98 & 0.77 \\
 $x$ & 0.52 & 0.24 & 0.64 & 0.61 \\
 $y$ & 0.16 & 0.20 & 0.92 & 0.57 \\
\hline\hline
\end{tabular}
\end{center}
\label{tab:Pvalues_multi_correct}
\end{table}

To explore the sensitivity of our test, we also generate simulated
data in which, for each data point, we introduce a mismatch between
the correlation angle of the reference distribution and that of the
distribution from which the sample is drawn. As previously, we draw
the correlation angle $\psi_\mathrm{ref}$ of the reference
distribution uniformly between $30^\circ$ and $60^\circ$, but we now
draw the sample from a distribution with a correlation angle
$\psi=\psi_\mathrm{ref}+\frac{3}{2}\left(\psi_\mathrm{ref} -45^\circ
\right)$. We again performed the test separately on three subsets of
the data for which the correlation angle $\psi_\mathrm{ref}$ of the
reference distribution lies in the ranges $[30^\circ,40^\circ]$,
$[40^\circ,50^\circ]$ and $[50^\circ,60^\circ]$, respectively. For
each of these subsets, it is easy to show that the mismatch angle
$\psi_\mathrm{mis}=\psi-\psi_\mathrm{ref}$ is distributed uniformly in
the ranges $[-22.5^\circ,-7.5^\circ]$, $[-7.5^\circ,7.5^\circ]$ and
$[7.5^\circ,22.5^\circ]$, respectively. Thus, the mean absolute
mismatch angle is $15^\circ$ for the first and third subsets, but zero
for the second subset.  The resulting $p$-values for these sub-sets
are listed in Table~\ref{tab:Pvalues_multi_wrong}, and follow a
similar pattern to the case $\psi_\mathrm{mis}=15^\circ$ in
Table~\ref{tab:Pvalues}, with the two-dimensional unable to reject the
null hypothesis, but the $x$- and $y$~marginal cases ruling out the
null hypothesis at extremely high significance for the first and third
subsets. For the second sub-set, for which the mean mismatch angle is zero,
the $x$- and $y$-marginal tests are consistent with the null hypothesis.

\begin{table}
\caption{As in Table~\ref{tab:Pvalues_multi_correct}, but for the case
  in which the correlation angle $\psi_\mathrm{ref}$ of the reference
  distribution is drawn uniformly between $30^\circ$ and $60^\circ$,
  and the corresponding data point is drawn from a distribution with a
  correlation angle
  $\psi=\psi_\mathrm{ref}+\frac{3}{2}\left(\psi_\mathrm{ref} -45^\circ
  \right)$.}
\begin{center}
\begin{tabular}{|l|c|c|c|}
\hline\hline
 & $\psi_\mathrm{ref} < 40^\circ$ & $40^\circ \le \psi_\mathrm{ref} < 50^\circ$ & $50^\circ \le \psi_\mathrm{ref}$ \\
 \hline
 2-D & 0.12 & 0.71 & 0.30 \\
 $x$ & $1.1 \times 10^{-6}$ & 0.49 & $ 1.9 \times 10^{-2}$ \\
 $y$  & $6.5 \times 10^{-6}$ & 0.90 & $ 1.3 \times 10^{-3}$ \\
\hline\hline
\end{tabular}
\end{center}
\label{tab:Pvalues_multi_wrong}
\end{table}
\subsection{Validation of Bayesian inference: \planck SZ clusters}
\label{sect:PlanckSZcase}

We now illustrate our method for the validation of posterior
distributions derived from Bayesian inference analyses. We apply this
approach to the subtle, real-world example of the estimation of
cluster parameters via their Sunyaev-Zel'dovich (SZ) effect in
microwave temperature maps of the sky produced by the \planck
mission. For a description of the \planck maps and the details of the
construction of the SZ likelihood, see
\cite{PlanckResultsSZ,nominal_planck_dr} and references therein.

\subsubsection{Cluster parameterization}

In this application, we are again considering a two-dimensional
problem, namely the validation of the derived posterior
distribution(s) for scale-radius parameter $\theta_\mathrm{s}$ and
integrated Comptonization parameter $Y_\mathrm{tot}$ of the galaxy
cluster(s). Two further parameters defining position of the cluster
centre are marginalised over, and several more model parameters are
held at fixed values with delta-function priors. The most important of
these fixed parameters are the those defining the
spherically-symmetric Generalised-Navarro-Frenk-White (GNFW) profile
used to model the variation of pressure with radius in the cluster
\citep{nagai_2007}.  Specifically, if $r_\mathrm{s}$ is the
scale-radius of the cluster, these parameters are
$(c_{500},\gamma,\alpha,\beta)$, where $c_{500}=r_{500}/r_\mathrm{s}$
with $r_{500}$ being the radius at which the mean density is 500 times
the critical density at the cluster redshift, and $\gamma$, $\alpha$
and $\beta$ describe the slopes of the pressure profile at $r \ll
r_\mathrm{s}$, $r \sim r_\mathrm{s}$ and $r > r_\mathrm{s}$,
respectively. The parameters are fixed to the values $(1.18, 0.31, 1.05, 5.49)$
 of the `universal pressure profile' (UPP) proposed by
\cite{arnaud_rexcess}.

\subsubsection{Evaluation of the posterior distributions}

In the analysis of \planck data, for each cluster the posterior
distribution of its parameters is obtained using the PowellSnakes
algorithm \citep{PwSII}. The marginalised
two-dimensional posterior in
$(\theta_\mathrm{s},Y_\mathrm{tot})$-space is evaluated on a
high-resolution rectangular grid. The posteriors derived are typically
complicated, non-Gaussian and vary between individual clusters, with
different levels of correlation, skewness, and flexion, as both the
signal and noise changes.

In order to achieve a tractable inference, the \planck SZ likelihood
used in these analyses necessarily makes a number of simplifying
assumptions. As described above, the cluster model assumes that the
pressure profile is well-described by a spherically-symmetric GNFW
profile, with the same fixed shape parameters for all clusters; this
may be a very poor approximation for some clusters. Aside from the
cluster model, the analysis assumes the background emission in which
the clusters are embedded is a realisation of a homogeneous Gaussian
process within the small patch of sky surrounding the cluster, whereas
in reality foreground signals from Galactic emission (mostly dust) and
infra-red compact sources add a non-Gaussian and anisotropic component
that dominates in certain areas of the sky. It is also assumed that
the cluster signal is convolved at each observing frequency with a
position-independent Gaussian beam, whereas in reality the effective
beams are not symmetric and vary across the sky.

\subsubsection{Validation of the posterior distributions}

Our method to validate the posteriors makes use of the fact that one
may generate simulations of the data that are often of (considerably)
greater sophistication and realism than can be modelled in the
inference process. In this application, the assumptions made by the \planck SZ
likelihood clearly divide into those related to the model of the
cluster itself and those associated with the background emission and
observing process. We will concentrate here on the latter set of
assumptions, but will return to the issue of the cluster model at the
end of this section. Of course, our validation process also
automatically provides a test of the software implementation of the
inference analysis.

We generate our simulations as follows.  The mass and redshift of the injected clusters are drawn from
a Tinker mass function \citep{tinker:2008} (assuming $h=0.7, \sigma_8= 0.8,
\Omega_{\text{m}}=0.3$).  The SZ parameters ($Y_{500},\theta_{500}$)
are calculated using a \planck\ derived scaling relation \citep{PlanckResultsSZ}.
 We inject the simulated clusters (each having the GNFW profile with the fixed UPP shape
 parameters assumed in the analysis), convolved with the appropriate effective beams, into the
real \planck sky maps. By doing so, we are thus testing all the
assumptions and approximations (both explicit and implicit) made in
the \planck SZ likelihood that are not associated with the cluster
model itself.    PwS is then run in detection mode to produce a catalogue of realistic
  detected clusters.  The injected values of the SZ observables ($Y_{tot},\theta_{s}$)
  are calculated from the injected ($Y_{500},\theta_{500}$) using the injected GNFW profile.

These simulations are then analysed using our inference pipeline, and
a posterior distribution in the
$(\theta_\mathrm{s},Y_\mathrm{tot})$-space is obtained for each
cluster. One subtlety that we must address is the issue of Eddington
bias, which arises from selecting clusters using a signal-to-noise
threshold. In the analysis of real data, this results in an incomplete
sample of the population of clusters at lower Compton-Y fluxes, since
only upward fluctuations are sampled.  This bias, although present
across the collection of posteriors derived, is independent of the
method and assumptions employed to obtain them. Here, we choose to
concentrate solely on the problem of parameter estimation in each
cluster, rather than the selection of the cluster
sample. Consequently, we ignore the issue of Eddington bias by
analysing the complete sample of injected clusters above a minimal
threshold of $Y_\mathrm{tot} > 3 \times 10^{-3}$ arcmin$^2$. This
results in a total catalogue of $N_\mathrm{cl}=918$ clusters.

\begin{figure*}
  \begin{center}
\includegraphics[height=6cm]{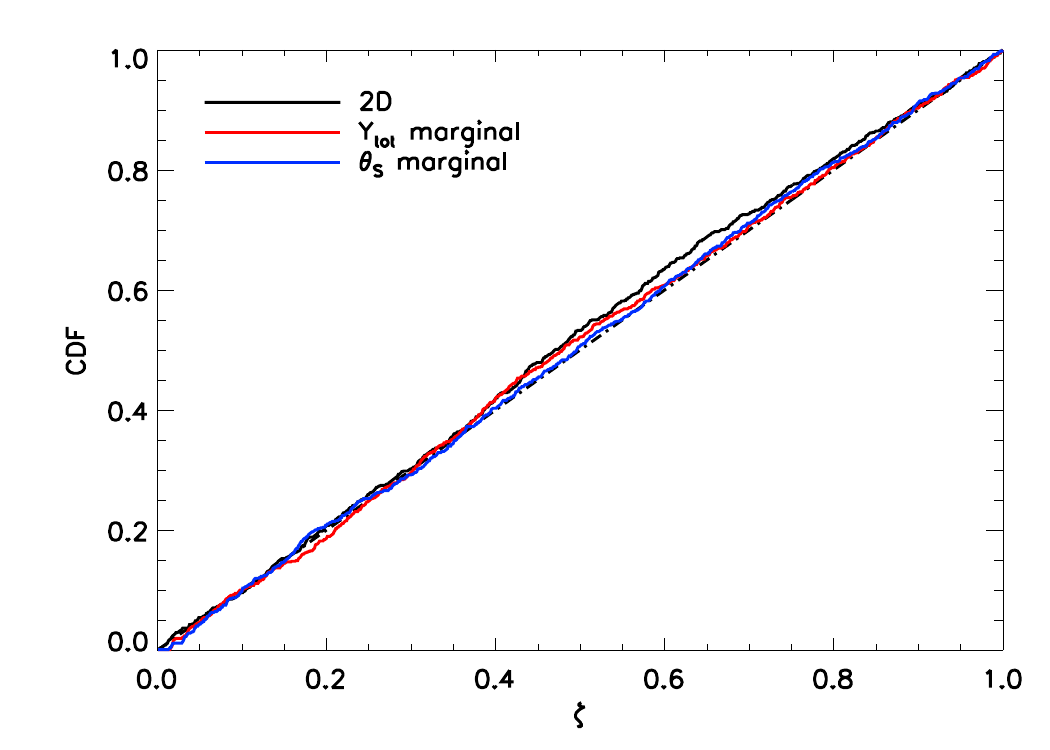}
\includegraphics[height=6cm]{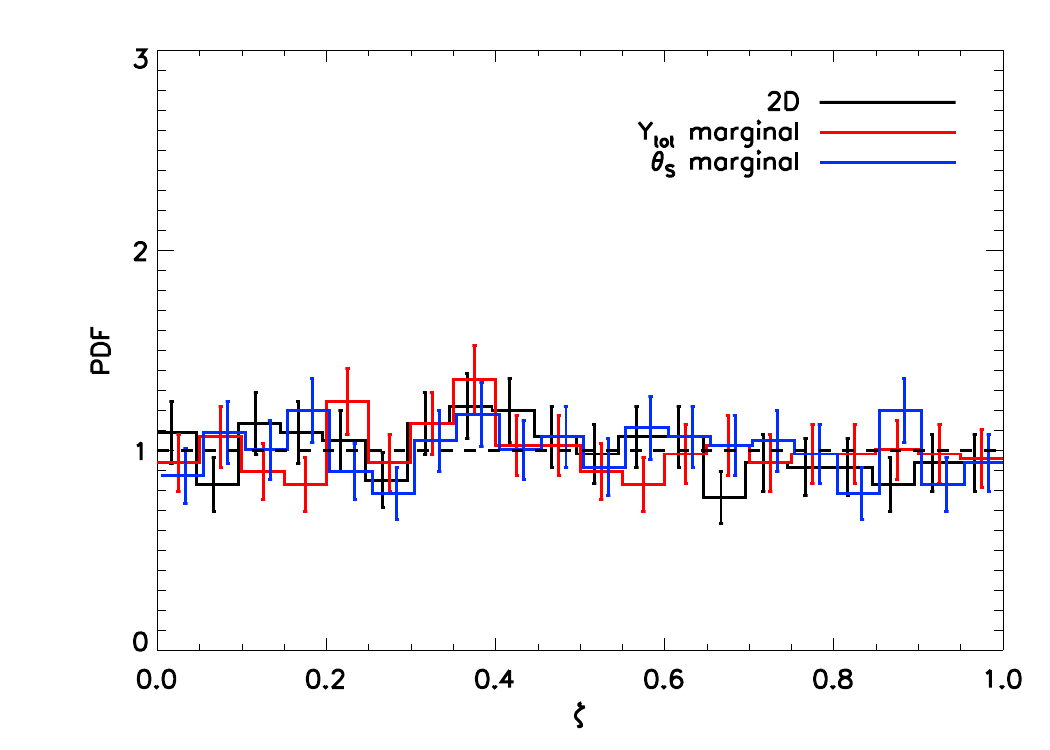}
    \caption{Left: the CDF of the standard uniform distribution
      (dot-dashed black line) and the empirical CDFs, for the full
      2-dimensional case (solid black line) and for the
      one-dimensional $Y_\mathrm{tot}$- and
      $\theta_\mathrm{s}$-marginals, respectively, (red and blue solid
      lines) of the $\{\zeta_k\}$-values obtained in the validation of
      posterior distributions
      $\mathcal{P}_k(\theta_\mathrm{s},Y_\mathrm{tot})$
      $(k=1,2,\ldots,N_\mathrm{cl})$ derived from $N_\mathrm{cl}=918$
      simulated clusters injected
      into real \planck sky maps. The injected clusters have the same
      fixed `shape' parameter values as assumed in the analysis,
      namely the `universal pressure profile'
      $(c_{500},\gamma,\alpha,\beta)=(1.18, 0.31, 1.05, 5.49)$.
      Right: the PDF of the standard uniform distribution (dashed
      black line) and the empirical PDFs of the corresponding
      $\{\zeta_k\}$-values (constructing by dividing them into 20
      bins) for the full 2-dimensional case (solid black line) and for
      the one-dimensional $Y_\mathrm{tot}$- and
      $\theta_\mathrm{s}$-marginals, respectively (red and blue solid
      lines); the error bars denote the Poissonian
      uncertainty.
    \label{fig:plancksz1}}
  \end{center}
\end{figure*}

The posterior distributions derived for each cluster,
$\mathcal{P}_k(\theta_\mathrm{s},Y_\mathrm{tot})$
$(k=1,2,\ldots,N_\mathrm{cl})$, are then validated using the method
described in Section~\ref{sec:validpost}. Thus, we test the null
hypothesis, simultaneously across all clusters, that the true
(injected) values $(\theta^{(k)}_\mathrm{s},Y^{(k)}_\mathrm{tot})$ are
drawn from the corresponding derived posterior
$\mathcal{P}_k(\theta_\mathrm{s},Y_\mathrm{tot})$. The empirical CDFs
(for the two-dimensional case and marginals) of the resulting
$\{\zeta_k\}$-values (one for each injected cluster), as defined in
(\ref{eqn:bayeszeta}), are shown in the left panel of
Fig.~\ref{fig:plancksz1} and the corresponding $p$-values obtained
from our test are given in the first column of
Table~\ref{tab:Pvalues_SZ}. One sees that all the empirical CDFs lie
close to that of a uniform distribution, although there is some
discrepancy visible for the two-dimensional case. This observation is
confirmed by the $p$-values, but all of them have $p> 0.05$, showing
that one cannot reject the null hypothesis even at the 5 per cent
level.

\begin{figure*}
  \begin{center}
\includegraphics[height=6cm]{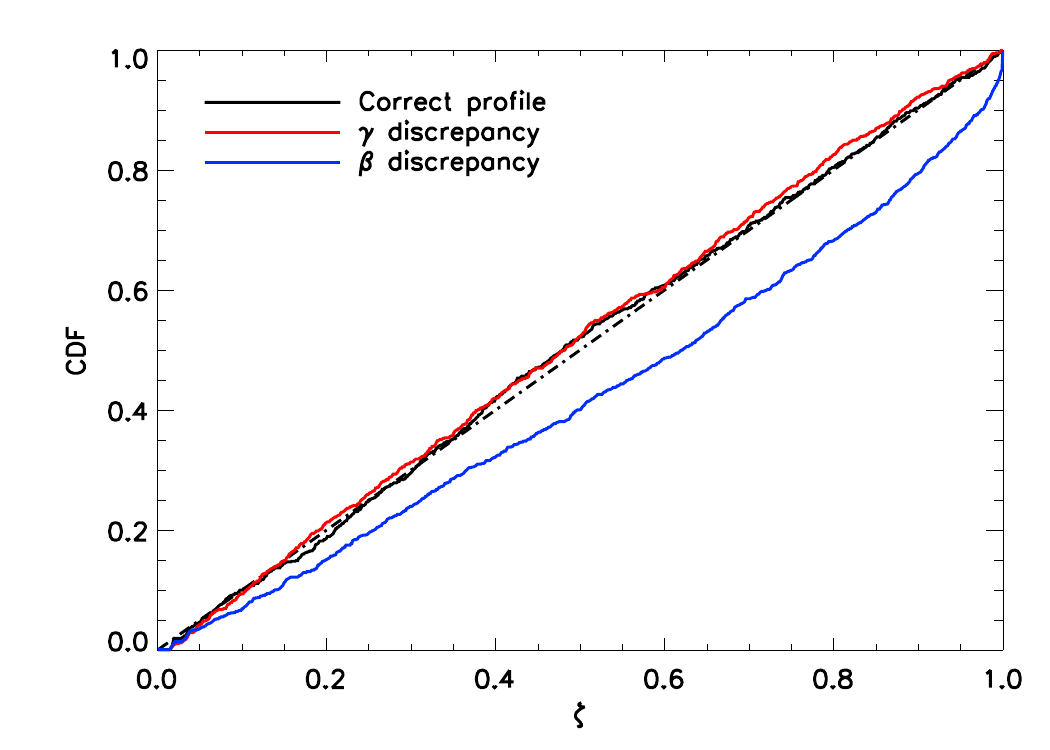}
\includegraphics[height=6cm]{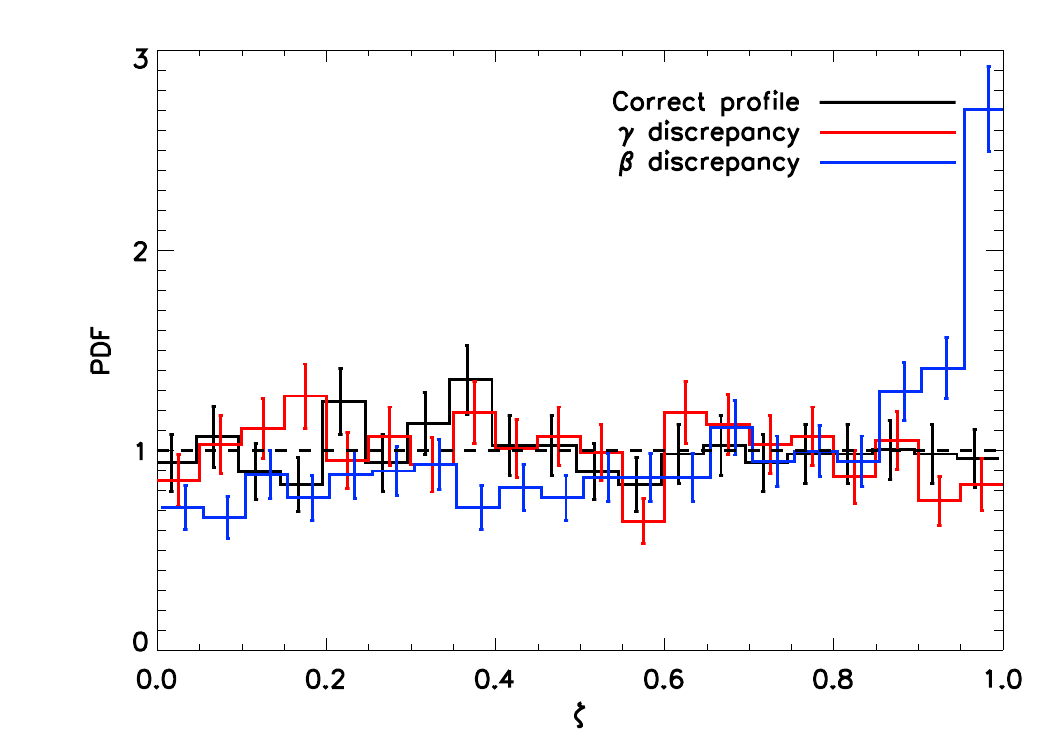}
    \caption{As in Fig.~\ref{fig:plancksz1}, but showing only results
      derived from the $Y_\mathrm{tot}$-marginal, for the case in
      which the injected clusters have $\gamma=0$ (red line) and
      $\beta=4.13$ (blue line), respectively.
    \label{fig:plancksz2}}
  \end{center}
\end{figure*}
\begin{table}
\caption{The $p$-values for the null hypothesis, as obtained from our
  test for validating Bayesian inference analyses applied to the
  posterior distributions
  $\mathcal{P}_k(\theta_\mathrm{s},Y_\mathrm{tot})$
  $(k=1,2,\ldots,N_\mathrm{cl})$ derived from $N_\mathrm{cl}=918$
  simulated clusters injected into
  real \planck sky maps. The first column gives the $p$-values for the
  case in which the injected clusters have the same fixed `shape'
  parameter values as assumed in the analysis, namely the `universal
  pressure profile' $(c_{500},\gamma,\alpha,\beta)=(1.16, 0.33, 1.06,
  5.48)$. The second and third columns correspond, respectively, to
  when the injected clusters have $\gamma=0$ and $\beta=4.13$; see
  text for details.}
\begin{center}
\begin{tabular}{l|c|c|c}
\hline\hline
 & Correct profile & $\gamma$ discrepancy & $\beta$ discrepancy \\
 \hline
2-D & 0.09 & $<10^{-8}$ & $<10^{-8}$ \\
$Y_\mathrm{tot}$ & 0.47 &  0.23 & $<10^{-8}$\\
$\theta_\mathrm{s}$ & 0.93 & 0.06 &$<10^{-8}$\\
\hline\hline
\end{tabular}
\end{center}
\label{tab:Pvalues_SZ}
\end{table}

For the purposes of interpretation, it is again useful to consider the
corresponding PDFs, which are plotted in the right panel of
Fig.~\ref{fig:plancksz1}. These are again constructed by dividing the
$\zeta$-range $[0,1]$ into 20 bins of equal width, which provides the
necessary resolution to constrain departures from uniformity in the
wings of the distribution ($\zeta>0.95$). We mentioned in
Section~\ref{sect:ndimKS} that the K-S test is not equally sensitive
to the entire interval $[0,1]$ and note here that one of its
insensitivities is to discrepancies in the wings.  Directly testing
whether the PDF in $\zeta$ is consistent with the uniform
distribution, particularly in at high $\zeta$-values, thus represents
a useful complement to the $n$-dimensional K-S test.  With a samples
size of $N_\mathrm{cl} = 918$ clusters and assuming Poisson statistics, one has a precision of
about 14 per cent on the PDF in each bin. We see from the figure that
the binned PDFs are all consistent with a uniform distribution.

We therefore conclude that the constraints on cluster parameters derived
from the \planck\ SZ likelihood are robust (at least to the sensitivity of our
test) to real world complications regarding foreground emission and beam
shapes.

\subsubsection{Mis-modelling of the cluster pressure profile}

Finally, we consider the robustness of our parameter inference to
mis-modelling the pressure profile of the clusters. To this end, we
generate two set of simulations in which the injected clusters have
different values of the GNFW shape parameters to those assumed in the
analysis. In the first set, we inject clusters with a core slope
$\gamma=0$, which is characteristic of morphologically-disturbed
clusters (\citealt{arnaud_rexcess}, \citealt{bolocam_profiles}). In
the second set, we assume $\beta=4.13$ for the outer slope, which is
the best-fit value for the mean profile of a sample of 62 bright SZ
clusters using a combination of \planck and X-ray data
\citep{Planck_PP}.

These two sets of simulations are then analysed and resulting
posterior distributions validated in the same way as discussed above.
For each set, the empirical CDF of the resulting $\{\zeta_k\}$-values,
as derived from the $Y_\mathrm{tot}$-marginal, are shown in the left
panel of Fig.~\ref{fig:plancksz2} and the $p$-values obtained from our
test (both for the two-dimensional case and the marginals) are given
in the second and third columns of Table~\ref{tab:Pvalues_SZ},
respectively. For a discrepancy in $\gamma$, one sees that the test
applied to the marginal posteriors is insensitive to mis-modelling the
cluster profile, but that the test on the joint two-dimensional
posteriors rejects the null hypothesis at extremely high
significance. By contrast, for a discrepancy in $\beta$, the test
applied to the two-dimensional posteriors and both marginals all
clearly reject the null hypothesis.

One may instead interpret the above results as indicating that the
inference process produces $Y_\mathrm{tot}$- and
$\theta_\mathrm{s}$-marginal posteriors (and parameter estimates
derived therefrom) that are robust (at least to the sensitivity of the
test) to assuming an incorrect value of $\gamma$ in the analysis, but
that their joint two-dimensional posterior is sensitive to this
difference. By contrast, the joint distribution and both marginals are
sensitive to assuming an incorrect value of $\beta$.

To assist in identifying the nature of this sensitivity, it is again
useful to consider the binned, empirical PDFs of the
$\{\zeta_k\}$-values; these are plotted in the right panel of
Fig.~\ref{fig:plancksz2} for the test applied to the
$Y_\mathrm{tot}$-marginal. In keeping with the other results, one sees
that the PDF is consistent with a uniform distribution in the case of
a discrepancy in $\gamma$, but is clearly different when there is a
discrepancy in $\beta$. Moreover, one sees that this difference is
most apparent at large $\zeta$-values. Indeed, the shape of the PDF is
consistent with an underestimation by $\sim 30$ per cent of the area
of the 1-$\sigma$ contour of the posterior distribution, which would
not be too serious in some applications.  Further investigation shows,
however, that the problem here is associated with the extent of the
derived posterior, but a bias low in the peak location of around 10
per cent. This bias is sub-dominant to the statistical uncertainty for
most clusters in the sample but is the dominant source of error for
high signal-to-noise clusters.  Furthermore, as a coherent systematic
across the whole sample, such an error could have damaging
influence on derived information such as scaling relations or
cosmological parameters.

\section{Conclusions}
\label{sect:Conclusions}
In this paper we firstly present a practical extension of the K-S test
to $n$-dimensions.
The extension is based on a new variable $\zeta$, that provides a
universal $\mathbb{R}^n \rightarrow [0,1]$ mapping based on the
probability content of the highest probability density
region of the reference distribution under consideration.
This mapping is universal in the sense that it is not distribution
dependent. By exploiting this property one may perform many
simultaneous tests, with different underlying distributions, and
provide an ensemble goodness-of-fit assessment.

We then present a new statistical procedure for validating Bayesian
posterior distributions of any shape or dimensionality.  The power of
the method lies in the capacity to test all of the assumptions in the
inferential machinery.  The approach goes beyond the testing of
software implementation and allows robust testing of the assumptions
inherent in the construction of the likelihood, the modelling of data
and in the choice of priors.
This approach enables the observables to inform our understanding of
the importance the various components of the model given the data.
It is, therefore, possible to tune the complexity of a model to
reflect the information actually available in the posterior, improving
the parsimony of the inference in keeping with Occam's Razor.
Conversely, important hidden parameters or overly-restrictive prior
assumptions can be identified and treated properly.

In the application to SZ cluster parameter inference from \planck
data, we demonstrate how the method can be applied to a large
population where the posterior distributions vary. The information
from a full cluster population can be combined to test the inference
in the presence of varying stochastic and systematic uncertainties, as
well as a varying signal component.  For this application, we have
found that the simplified \planck cluster likelihood is robust to real world
complications such as Galactic foreground contamination and realistic
beams.  The sensitivity of the inference to prior assumptions on the
outer slope of the pressure profile has been identified, as has the
insensitivity to assumptions on the pressure profile in the core
regions of the cluster for the inference of the integrated Compton-Y
parameter.

This approach could be of use in improving the pool of available SZ
data from high-resolution microwave experiments, which to date have
provided either non-Bayesian point estimates for cluster parameters or
parameter proxies (\citealt{act}, \citealt{spt}), or unvalidated
Bayesian posterior distributions \citep{ami}.
These experiments have different dependencies on cluster parameters
given their different resolutions, parameterisations and observation
frequencies.
A fuller understanding of the nature of these dependencies and the
sensitivity of the derived posteriors to assumptions of the cluster
model will ensure the robustness of the results and maximise the wider
scientific returns due to the complementarity of the data-sets.

Beyond astronomy, the methodology we have introduced may be applied to
Bayesian inference more generally, in any situation where higher
levels of complexity and fidelity can be introduced into simulations
than can be allowed for in a tractable analysis, or where there exists
a pool of pre-existing real data with known outcomes.

\section{Acknowledgements}
This work was supported by the UK Space Agency under grant
ST/K003674/1.
This work was performed using the Darwin Supercomputer of the
University of Cambridge High Performance Computing Service
(http://www.hpc.cam.ac.uk/), provided by Dell Inc. using Strategic
Research Infrastructure Funding from the Higher Education Funding
Council for England and funding from the Science and Technology
Facilities Council.
The authors thank Mark Ashdown, Steven Gratton and Torsten En\ss{}lin
for useful discussions, and Hardip Sanghera and the Cambridge HPC
Service for invaluable computing support. 
We also thank the referee, 
John Peacock, for numerous insightful comments that undoubtedly
improved the paper.

\bibliographystyle{mn2e}
\setlength{\bibhang}{2.0em}
\setlength\labelwidth{0.0em}
\bibliography{references} 

\appendix
\section{Some properties of HPD regions}
\label{sec:HPDprops}

\subsection{HPD region defines the smallest `error bar'}
\label{sec:ApxHPDerrorBar}
We wish to show that HPD$_\zeta$ is the region of smallest volume that
contains an integrated probability $\zeta$. Suppose one begins with
HPD$_\zeta$ and attempts to swap a small region of volume $d^n
\bmath{x}_{\scriptsize \mbox{HPD}}$ from the interior of the HPD with
a small region of volume $d^n \bmath{x}_{{\widetilde{\scriptsize
      \mbox{HPD}}}}$ from outside the HPD. For the total
probability contained to remain unchanged, one requires
\begin{equation}
d\zeta_{\scriptsize \mbox{HPD}} = d\zeta_{\widetilde{\scriptsize \mbox{HPD}}},
\label{eq:HPDVolMin_00}
\end{equation}
which, using equation (\ref{eq:deltaZ_Empirical}), leads to
\begin{equation}
f(\bmath{x}_{\widetilde{\scriptsize \mbox{HPD}}}) =
f(\bmath{x}_{\scriptsize \mbox{HPD}})\,\left(\frac{d^n
  \bmath{x}_{\scriptsize \mbox{HPD}}}{d^n
  \bmath{x}_{{\widetilde{\scriptsize \mbox{HPD}}}}}\right).
\label{eq:HPDVolMin_01}
\end{equation}
In order for the new region to have a smaller volume, the quantity in
parentheses, ${d^n \bmath{x}_{\scriptsize \mbox{HPD}}} / {d^n
  \bmath{x}_{{\widetilde{\scriptsize \mbox{HPD}}}}}$, must be larger
than unity. This would imply that
\begin{equation}
f(\bmath{x}_{\widetilde{\scriptsize \mbox{HPD}}}) > f(\bmath{x}_{\mbox{\scriptsize HPD}}),
\label{eq:HPDVolMin_02}
\end{equation}
which contradicts the definition of HPD, see
Section~\ref{sect:Hpds}. Therefore, for a given probability content
$\zeta$, the HPD encloses the region of smallest volume.

\subsection{HPD probability content is uniformly distributed}
\label{sec:ApxHPDisUniform}

Suppose some point $\bmath{x}_\ast$ in an $n$-dimensional parameter
space is drawn from the PDF $f(\bmath{x})$ and the corresponding
probability content of the HPD whose boundary passes through
$\bmath{x}_\ast$ is $\zeta_\ast$. The probability distribution of
$\zeta_\ast$ may be written (somewhat baroquely) as
\begin{eqnarray}
g(\zeta_\ast) & = & \partial_{\zeta_\ast}\int_{-\infty}^{\zeta_\ast}
d\xi\, g(\xi), \\
& = &
\partial_{\zeta_\ast}\int_{-\infty}^{\zeta_\ast} d\xi \int
d^n\bmath{x}\,f(\bmath{x})\, \delta(\zeta(x)-\xi),
\end{eqnarray}
where $\zeta(\bmath{x})$ is the mapping defined in
(\ref{eq:HPD_varDef}) and the integration on $\bmath{x}$ extends over
the full $n$-dimensional parameter space. Performing the integral on $\xi$,
one obtains
\begin{equation}
g(\zeta_\ast) = \partial_{\zeta_\ast} \int
d^n\bmath{x}\,f(\bmath{x})\, \Phi(\zeta_\ast-\zeta(x)),
\end{equation}
where $\Phi$ is the Heaviside step function. The integral on the RHS
is, however, equal simply to $\zeta_\ast$. Thus, as required, one obtains
$g(\zeta_\ast)=1$ within the allowed range $\zeta_\ast \in [0,1]$.

\label{lastpage}
\end{document}